\documentclass[11pt]{article}

\newif\ifnotes
\notesfalse

\usepackage{hyperref}
\usepackage{fullpage}
\usepackage{amssymb}
\usepackage{amsmath}
\usepackage{amsthm}
\usepackage{amsfonts}
\usepackage{bm}
\usepackage{enumitem}
\usepackage{color}
\usepackage{comment}
\usepackage[capitalize]{cleveref}
\usepackage[dvipsnames]{xcolor}
\usepackage{float}
\usepackage[T1]{fontenc}
\usepackage{todonotes}
\usepackage{asymptote}
\usepackage{mdframed}
\usepackage[most]{tcolorbox}
\usepackage{hyperref}
\usepackage{enumitem}
\usepackage{framed}
\usepackage{mdframed}
\usepackage{scrextend}
\usepackage{multirow}
\usepackage{ifthen}
\usepackage{bbm}

\usepackage[T1]{fontenc}

\definecolor{denim}{rgb}{0.08, 0.38, 0.74}
\definecolor{periwinkle}{rgb}{0.6, 0.6, 0.95}
\definecolor{wildblueyonder}{rgb}{0.64, 0.68, 0.82}
\definecolor{wisteria}{rgb}{0.91, 0.72, 1.00}
\definecolor{thistle}{rgb}{0.85, 0.75, 0.85}
\definecolor{byzantium}{rgb}{0.44, 0.16, 0.39}
\definecolor{deeplilac}{rgb}{0.6, 0.33, 0.73}
\definecolor{jazzberryjam}{rgb}{0.55, 0.04, 0.37}

\usepackage{hyperref}
\hypersetup{
    colorlinks=true,
    linkcolor=jazzberryjam,
    filecolor=jazzberryjam,
    citecolor=deeplilac,
    urlcolor=deeplilac,
}
\usepackage[hyperpageref]{backref}

\newtheorem{theorem}{Theorem}[section]
\newtheorem{lemma}[theorem]{Lemma}
\newtheorem{claim}[theorem]{Claim}

\theoremstyle{definition}
\newtheorem{definition}[theorem]{Definition}
\newtheorem{remark}[theorem]{Remark}

\Crefname{theorem}{Theorem}{Theorems}
\Crefname{claim}{Claim}{Claims}
\Crefname{lemma}{Lemma}{Lemmas}
\Crefname{proposition}{Proposition}{Propositions}
\Crefname{corollary}{Corollary}{Corollaries}
\Crefname{definition}{Definition}{Definitions}

\newcommand{\ECC}{\mathsf{ECC}}

\newcommand{\DEC}{\mathsf{DEC}}

\newcommand{\op}{\mathsf{op}}

\newcommand{\rewind}{\,\leftarrow}

\newcommand{\speaker}{\mathsf{speaker}}
\newcommand{\listener}{\mathsf{listener}}
\newcommand{\passer}{\mathsf{passer}}

\newcommand{\maj}{\text{maj}}

\newcommand{\bbN}{\mathbb{N}}

\newcommand{\bbZ}{\mathbb{Z}}
\newcommand{\bbE}{\mathbb{E}}

\makeatletter
\newcommand{\customlabel}[2]{%
   \protected@write \@auxout {}{\string \newlabel {#1}{{#2}{\thepage}{#2}{#1}{}} }%
   \hypertarget{#1}{#2}
}
\makeatother

\newcommand{\protocol}[3]{
    \stepcounter{figure}
    \vspace{0.15cm}
    { \small
    \begin{tcolorbox}[breakable, enhanced, colback=wisteria!10]
    \begin{center}
    {\bf \underline{Protocol~\customlabel{prot:#2}{\thefigure}: #1}}
    \end{center}
    
    #3
    \end{tcolorbox}
    }
}

\newcommand\numberthis{\addtocounter{equation}{1}\tag{\theequation}}

\newcounter{casenum}
\newenvironment{caseof}{\setcounter{casenum}{1}}{\vskip.5\baselineskip}
\newcommand{\case}[2]{
    \ifthenelse{\equal{\value{casenum}}{0}}{
    \vskip.5\baselineskip\par\noindent
    }{}
    {\it Case \arabic{casenum}:} {\it #1}
    \vskip0.1\baselineskip
    \begin{addmargin}[1.5em]{1em}
    #2
    \end{addmargin}
    \addtocounter{casenum}{1}
}

\newcounter{subcasenum}
\newenvironment{subcaseof}{\setcounter{subcasenum}{1}}{\vskip.5\baselineskip}
\newcommand{\subcase}[2]{
    \vskip.5\baselineskip\par\noindent 
    {\it Subcase \arabic{casenum}.\arabic{subcasenum}:} {\it #1} \vskip0.1\baselineskip
    \begin{addmargin}[1.5em]{1em}
    #2
    \end{addmargin}
    \addtocounter{subcasenum}{1}
}

\begin{document}

\title{The Optimal Error Resilience of Interactive Communication Over Binary Channels}
\author{Meghal Gupta\thanks{Email: \texttt{meghal@mit.edu}}\\Microsoft Research 
\and Rachel Yun Zhang\thanks{Email: \texttt{rachelyz@mit.edu}}\\Massachusetts Institute of Technology}
\date{\today}

\sloppy
\maketitle
\begin{abstract}
In interactive coding, Alice and Bob wish to compute some function $f$ of their individual private inputs $x$ and $y$. They do this by engaging in a non-adaptive (fixed order, fixed length) interactive protocol to jointly compute $f(x,y)$. The goal is to do this in an error-resilient way, such that even given some fraction of adversarial corruptions to the protocol, both parties still learn $f(x,y)$. 

In this work, we study the optimal error resilience of such a protocol in the face of adversarial bit flip or erasures. While the optimal error resilience of such a protocol over a large alphabet is well understood, the situation over the \emph{binary} alphabet has remained open. In this work, we resolve this problem of determining the optimal error resilience over binary channels. In particular, we construct protocols achieving $\frac16$ error resilience over the binary bit flip channel and $\frac12$ error resilience over the binary erasure channel, for both of which matching upper bounds are known. We remark that the communication complexity of our binary bit flip protocol is polynomial in the size of the inputs, and the communication complexity of our binary erasure protocol is linear in the size of the minimal noiseless protocol computing $f$.
\end{abstract}
\thispagestyle{empty}
\newpage
\tableofcontents
\pagenumbering{roman}
\newpage
\pagenumbering{arabic}

\section{Introduction}

Interactive coding is an interactive analogue of error correcting codes \cite{Shannon48, Hamming50}, that was introduced in the  seminal work of Schulman~\cite{Schulman92,Schulman93,Schulman96} and has been an active area of study since.
While error correcting codes address the problem of sending a {\em message} in a way that is resilient to error, interactive coding addresses the problem of   converting an {\em interactive protocol} to an error resilient one.

Suppose two parties, Alice and Bob, each with a private input, engage in a protocol $\pi_0$ to jointly
compute a function $f$ of their private inputs. \emph{Given such a protocol $\pi_0$, can we design a protocol $\pi$ that computes $f$ and at the same time is resilient to adversarial errors?} Schulman \cite{Schulman96} answered the question in the affirmative, presenting a scheme that is resilient to $\frac{1}{240}$\footnote{Whenever we say that a protocol has resilience $r \in [0, 1]$ in the introduction and overview, we mean that for any $\epsilon$, there exists an instantiation that achieves resilience $r - \epsilon$.} adversarial corruptions (bit flips) over a binary channel with a constant information rate.\footnote{Constant information rate means that the error-resilient protocol incurs only a constant multiplicative overhead
to the communication complexity.} This work begs the natural question: \emph{what is the maximum error resilience possible?} This is precisely the focus of our work.


Two natural types of corruption to consider are \emph{bit flip} (where the adversary can replace a symbol with one of their choice) and \emph{erasure} (where the adversary can replace a symbol with $\perp$).
In both of these settings, there are known protocols that achieve optimal error resilience for large constant-sized alphabets. In the bit flip setting, Braverman and Rao~\cite{BravermanR11} constructed a protocol which achieves the optimal error resilience of $\frac14$.  In the erasure setting, 
\cite{FranklinGOS15,EfremenkoGH16} constructed protocols achieving the optimal error resilience $\frac12$. Corresponding impossibility bounds are known \cite{BravermanR11,FranklinGOS15}.

However, despite much effort, the optimal error resilience for both types of corruption over a \emph{binary} alphabet is still unknown. For both types of corruption, a protocol achieving error resilience of $r$ over a large alphabet trivially translates to a protocol over a binary alphabet with error resilience of $\frac{r}2$ by replacing every letter of the large alphabet with a binary error correcting code of relative distance $\frac12$. Thus, the results of~\cite{BravermanR11,FranklinGOS15} give protocols that achieve error resilience $\frac18$ over the binary bit flip channel and $\frac14$ over the binary erasure channel. 

Unfortunately, the corresponding impossibility bounds \cite{BravermanR11, FranklinGOS15} for large alphabets do not lead to matching impossibility bounds in the binary case. Over the binary bit flip channel, the best known impossibility bound is $\frac16$ \cite{EfremenkoGH16}, and over the binary erasure channel, the best known impossibility bound is $\frac12$ \cite{FranklinGOS15}. Pinning down the exact constant between $\frac18$ and $\frac16$, and between $\frac14$ and $\frac12$, has been an intriguing open problem. 

There has been some recent work towards this goal. Over the binary bit flip channel, \cite{EfremenkoKS20b} broke the $\frac18$ barrier for the first time, describing a protocol that achieves $\frac5{39}$ resilience to adversarial bit flips. Over the binary erasure channel, \cite{EfremenkoGH16} gives a protocol achieving a $\frac13$ resilience to erasures. Nonetheless, the exact values of the optimal error resilience over the binary bit flip and erasure channels have remained unknown since their initial active investigation by \cite{BravermanR11} in 2011 and \cite{FranklinGOS15} in 2013.

In this work, we resolve the question of the optimal error resilience of a non-adaptive (fixed order, fixed length) protocol over a binary alphabet. Specifically, we show that the known impossibility bounds are tight: \emph{we construct protocols achieving error resilience $\frac16$ to adversarial bit flips and $\frac12$ to adversarial erasures.}

\subsection{Our Results}







We show the following result for two-party communication over the binary bit flip channel.

\begin{theorem}
    For any function $f$, there exists a non-adaptive binary interactive protocol computing the function $f(x, y)$ that is resilient to $\frac16 - \epsilon$ adversarial bit flips with probability $1 - 2\exp(-\Omega(\epsilon n))$. For inputs of size $n$, the communication complexity is $O_\epsilon(n^2)$ and the runtime of the parties is $C(\epsilon) \cdot n^{O(1)}$ for some constant $C(\epsilon)$.
\end{theorem}


Until our result, it was only known how to achieve $\frac16$ error resilience if Alice and Bob are given the extra power to know, instantly, what the other party received at the other end of the channel when they send a message. This additional power is known as \emph{feedback}~\cite{EfremenkoGH16,GellesH17,Berlekamp64,Berlekamp68,Zigangirov76,SpencerW92} and is given at no cost. It is thought to give considerably more power to the parties, as there is never any uncertainty about what the other party has heard so far. An error resilience of $\frac16$ is known to be tight even in this model \cite{EfremenkoGH16}, but all protocol constructions rely crucially on the fact that the sender can always send specifically the piece of information about their input that the receiver needs to hear.

Our $\frac16$ error resilient protocol shows that protocols \emph{without feedback} can do just as well. The surprising implication is that the ability to know what messages are received by the other party actually grants no additional power!

\begin{remark} 
    The communication complexity and runtime of our protocol are polynomial in the size of Alice and Bob's inputs, rather than on the minimal size of an error-free protocol $\pi_0$ computing $f$, which can be exponentially smaller. This may result in an exponential blowup in communication complexity relative to a corresponding error-free protocol $\pi_0$ for simpler functions Alice and Bob might want to compute. We leave it as an open problem to construct (or disprove the existence of) a protocol with $\left(\frac16-\epsilon\right)$-error resilience whose communication complexity and runtime are polynomial in $|\pi_0|$ or ideally even linear.
\end{remark}

\noindent
Over the binary erasure channel, we show the following efficient result:

\begin{theorem}
    For any interactive binary protocol $\pi_0$ computing a function $f(x,y)$ of Alice and Bob's inputs $x,y\in\{0,1\}^n$, there exists a non-adaptive interactive binary protocol $\pi$ computing $f(x, y)$ that is resilient to $\frac12 - \epsilon$ adversarial erasures. The communication complexity and runtime for each party are both $O_\epsilon(|\pi_0|)$.
\end{theorem}

We note that $\frac12$ is in fact the maximal possible erasure resilience of a two-party interactive protocol over \emph{any} alphabet, as the adversary can simply erase all messages of the party that speaks less. 
Previous work \cite{FranklinGOS15,EfremenkoGH16} constructed protocols resilient to $\frac12$ erasures over larger alphabets. 
In our work, we show that the alphabet size makes no difference to the optimal erasure resilience. This contrasts with the bit flip model, where an error resilience of $\frac14$ is attainable over large alphabets, while over the binary alphabet it is capped at $\frac16$.

\section{Technical Overview}\label{sec:overview}

We now give an overview of our two protocols achieving maximal error resilience over the binary bit flip channel and binary erasure channel respectively. 

\subsection{Our Binary Protocol Resilient to $\frac16$ Bit Flips} \label{sec:overview-1/6}


We begin with the following (flawed) approach, which achieves an error resilience of $\frac18$: Alice sends an error correcting code $\ECC(x)$, and Bob sends $\ECC(y)$. Since $\ECC$ has relative distance of at most $\frac12$, the adversary can simply flip $\frac14$ of the bits of the party that speaks less so that the other party cannot distinguish between two inputs.

Recall that the maximum error resilience of any binary two-party protocol is at most $\frac16$ of the total communication, or $\frac13$ of either party's communication. The above flawed protocol only had error resilience of $\frac14$ of either party's communication. In order to increase this to $\frac13$ for one of the parties, we introduce a new \emph{question-answer} approach: in each round of interaction, Alice asks a question (encoded with an $\ECC$) about Bob's input, and then Bob responds with one of \emph{four} answers. More specifically,
\begin{itemize}
    \item Alice tracks a guess $\hat{y}$ for Bob's input $y$ initially set to $\emptyset$, and a counter $c_A$ indicating her confidence for $\hat{y}$ initially set to $0$. Each round, she sends $\hat{y}$ encoded in an $\ECC$ to Bob as her question.
    \item Bob responds with one of four operations to do to $\hat{y}$ as his answer: append 0 ($0$), append 1 ($1$), delete the last bit ($\rewind$), or ``bingo -- you got it right!'' ($\bullet$). 
    \item Alice updates based on the answer as follows: if she receives $\bullet$, she increases $c_A$ by $1$ since Bob is informing her that her guess is correct. If she receives $0$, $1$, or $\rewind$ and if $c_A=0$, she makes the corresponding adjustment to the string $\hat{y}$ (append $0$ or $1$, or delete the last bit). Otherwise if $c_A\neq 0$, she simply decreases $c_A$ by $1$ without making the corresponding adjustment to $\hat{y}$.
\end{itemize}
Ultimately, as long as Alice receives Bob's correct answer at least $|y|$ more times than she receives a wrong answer, she will output the correct answer. The key point is that these four responses from Bob can have distance $\frac23$ (e.g., $000$, $110$, $011$, $101$). Now, if the adversary simply corrupts Bob's answers, she'd have to corrupt $\approx \frac12 \cdot \frac23 = \frac13$ of Bob's rounds. 

While this achieves $\frac13$ resilience for Bob's communication, Alice's communication is still only $\frac14$-error resilient as she is sending an error correcting code. In order to attain $\frac13$ error resilience for \emph{both} parties, both parties will need to \emph{simultaneously ask and answer a question} each round. Specifically, 
\begin{itemize}
\item Alice and Bob each track a guess $\hat{y}$ or $\hat{x}$ respectively for the other party's input, as well as a counter $c_A$ or $c_B$. 
\item Each round, Alice sends $\ECC(\hat{y}, x^*, \delta)$: $\hat{y}$ is her question, $x^*$ is the question she just heard from Bob and $\delta$ is the instruction that brings $x^*$ one character closer to her input $x$ (or $\bullet$ if $x^*=x$). Similarly, Bob sends $\ECC(\hat{x}, y^*, \delta)$. 
\item When Alice receives the message $\ECC(x^*, y^*, \delta^*)$\footnote{Alice may also receive a message that is only partially corrupted (i.e., does not correspond to a codeword of the form $\ECC(x^*, y^*, \delta^*)$). We address partial corruptions later, and for now assume that the adversary must always corrupt a message to another valid codeword.} from Bob, she updates $\hat{y}$ according to the instruction $\delta^*$, but only if $y^* = \hat{y}$ (intuitively, because she should not update if Bob is answering the wrong question). Bob does the same.
\end{itemize}

The $\ECC$ we use has relative distance $\ge \frac12$ between all pairs of codewords, and $\ge \frac23$ for pairs of codewords of the form $\ECC(x', y', \delta_0)$ and $\ECC(x', y', \delta_1)$ with $\delta_0 \not= \delta_1$ (or equivalently $\ECC(y', x', \delta_0)$ and $\ECC(y', x', \delta_1)$ with $\delta_0 \not= \delta_1$). (We construct such an $\ECC$ explicitly in Claim~\ref{claim:ecc}.) Now if the adversary only corrupts Bob's answer, specifically only corrupting $\delta$, it will require $\approx \frac12\cdot \frac23=\frac13$ corruptions of Bob's communication for Alice to output the wrong $y$.

However, it seems as if we have lost the advantage we gained earlier: the adversary can corrupt both Bob's answer $\delta$ \emph{and} his question $\hat{x}$ in half the rounds so that Alice receives $\ECC(x', \hat{y}, \delta')$, where $x' \not= \hat{x}$ and $\delta' \not= \delta$. This attack every other round only requires corrupting $\frac12$ rather than $\frac23$ of Bob's message had they only corrupted $\delta$ and not $\hat{x}$ as well. Alice still does not make progress over time, and the adversary only needs to corrupt $\approx \frac12\cdot \frac12=\frac14$ of Bob's total communication.

Let us analyze this situation more closely. We count progress as the number of good updates (getting the guess $\hat{y}$ or $\hat{x}$ closer to the other party's input or adjusting $c_A$ or $c_B$ correctly) minus the number of bad updates (getting $\hat{y}$ or $\hat{x}$ further from the other party's input or adjusting $c_A$ or $c_B$ incorrectly). The adversary can corrupt $\frac12$ of the bits in Bob's message to get $-1$ progress for Alice and $0$ progress for Bob (since $x' \not= \hat{x}$, Alice answers the wrong question for Bob so that he performs no update), from an original progress count of $(+1, +1)$ had no corruptions occurred. So, when the adversary corrupts half of Bob's messages, Alice's final progress is $0$, so that she does not know $y$ at the end of the protocol. 
However, Bob's progress still increases at a steady rate! If there were a way to exploit this, so that when Bob has made a lot of progress (i.e., $\hat{x}=x$) it becomes harder to cause negative progress for Alice, perhaps we could achieve our goal of $\frac13$ error resilience for each party.

To do this, we make the following probabilistic change to the way Alice (likewise Bob) makes updates to her current guess $\hat{y}$ upon receiving $\ECC(x^*,\hat{y},\delta^*)$: 
\begin{itemize}
    \item Alice only updates $\hat{y}$ with probability $1$ if the question she just received is equal to her input (i.e., $x^*=x$), otherwise (if $x^*\neq x$) she updates $\hat{y}$ with probability only $0.5$.
\end{itemize} 
This way, when Bob has made lots of progress so that $\hat{x} = x$, if the adversary corrupts Bob's message to be $\ECC(x^* \not= x, \hat{y}, \delta^*)$, Alice only updates $\hat{y}$ with probability $0.5$ --- this is $-0.5$ progress instead of $-1$! Now, the adversary has to corrupt two out of every three of Bob's messages in order for Alice to remain at $0$ progress, bringing us up to a $\frac13$ corruption rate for Bob's messages. The key idea is that corrupting both Bob's question $\hat{x}$ and instruction $\delta$, which only involves corrupting $\frac12$ of the message rather than $\frac23$, becomes less damaging to Alice's progress than corrupting only $\delta$, which requires $\frac23$ corruption. This only works when Bob knows $x$ (so $\hat{x} = x$), but this is exactly what we wanted --- we needed to prevent the adversary from being able to cheaply keep Alice from making progress when Bob has made a lot of progress.

This change in update probability introduces a new situation, which is that when both Alice and Bob have made \emph{little} progress ($\hat{y} \not= y$ and $\hat{x} \not= x$), the adversary can corrupt Bob's message to be $\ECC(x \not= \hat{x}, \hat{y}, \delta')$ so that Alice performs a bad update with probability $1$, i.e. $(-1, 0)$ progress, from an original $(+0.5, +0.5)$ progress without corruption. Then, if the adversary corrupts one message every two rounds of interaction, the total progress between Alice and Bob remains $0$! To remedy this, we have Bob (and similarly Alice) perform an additional update:
\begin{itemize}
    \item When Bob receives a message of the form $\ECC(y^*,x^*\neq \hat{x}, \bullet$), he brings his current guess $\hat{x}$ one character closer to $x^*$ (or adjusts $c_B$ by $1$) with probability 0.5.
\end{itemize} 
Now, when Bob receives Alice's response $\ECC(\hat{y}, x, \bullet)$ to his corrupted question, he performs a $+0.5$ update, so that the total effect of the adversary's corruption to Bob's message is $(-1, +0.5)$, i.e. $-0.5$ collective progress. This makes it so that when $\hat{y} \not= y$ and $\hat{x} \not= x$, the adversary must corrupt on average $\frac13$ of a party's messages, or $\frac16$ the total communication, in order to prevent collective progress from being made. (Then, when a lot of collective progress has been made, at least one of the parties, say Bob, must have $\hat{x} = x$, so as discussed above the adversary must corrupt at least $\frac16$ of the communication to prevent Alice from also making progress.)

\paragraph{Our protocol.}
To summarize, here is an outline of our protocol from Alice's perspective, assuming that she always receives a full codeword. She holds a guess $\hat{y}$ and a confidence counter $c_A$, initialized to $\emptyset$ and $0$ respectively. Each round, she does the following: 
\begin{itemize}
    \item Alice receives $\ECC(x^*,y^*,\delta^*)$ from Bob.
    \item She updates $\hat{y}$ and $c_A$ as follows:
    \begin{itemize}
        \item If $y^*=\hat{y}$ and $x^* = x$, she updates $(\hat{y},c_A)$ according to $\delta^*$ with probability $1$.
        \item If $y^*=\hat{y}$ and $x^* \not= x$, she updates $(\hat{y},c_A)$ according to $\delta^*$ with probability $0.5$.
        \item If none of the first two conditions hold, meaning that $y^* \not= \hat{y}$, and if $\delta^*=\bullet$, she does the following update with probability $0.5$: she decreases $c_A$ by $1$ if $c_A>0$ and otherwise brings $\hat{y}$ one step closer to $y^*$.
    \end{itemize}
    \item Finally, she computes $\delta$ which brings $x^*$ closer to $x$ (or $\bullet$ if $x^*=x$) and sends Bob $\ECC(\hat{y},x^*,\delta)$.
\end{itemize}

\paragraph{Dealing with partial corruptions.}
It turns out this protocol works as long as Alice and Bob always receive a (possibly incorrect) codeword. We are almost done, but we need to specify their behaviors when they receive partially corrupted messages. To do this, we say that when Alice receives a message from Bob, she ``rounds'' to the nearest full codeword and does the corresponding update with some lower probability depending on the distance to the codeword. Precisely:
\begin{itemize}
    \item Alice rounds to a codeword $\ECC(x, \hat{y}, \delta)$ if the received message is relative distance $d^* < \frac13$ away, and otherwise she rounds to a codeword $\ECC(x^*, y^*, \delta^*)$ if the relative distance $d^*$ is $< \frac16$. (At most one such rounded codeword can exist since the relative distance between any two codewords is $\ge \frac12$ and between two codewords $\ECC(x, \hat{y}, \delta_0)$ and $\ECC(x, \hat{y}, \delta_1)$ it is $\frac23$.) 
    \item She performs the corresponding update with probability $1 - 3d^*$ or $0.5 - 3d^*$ respectively and then replies with $\hat{y}$, the value of Bob's question $x$ or $x^*$ in the rounded codeword, and an instruction $\delta$ on how to update it, all jointly encoded with $\ECC$. If no rounded codeword exists, she does no update and sends a message of the form $\ECC(\hat{y}, x, \bullet)$. 
\end{itemize}
For a formal description of the protocol and a detailed analysis, we refer the reader to Section~\ref{sec:toggle}.


\subsection{Our Binary Protocol Resilient to $\frac12$ Erasures} \label{sec:overview-1/2}

We construct a protocol that achieves an $\frac12 - \epsilon$ erasure resilience for any $\epsilon > 0$. In our construction, we leverage a key property of the erasure channel: \emph{communication can be delayed but not wrong}. To recreate the transcript of a (w.l.o.g. alternating) noiseless protocol $\pi_0$, Alice and Bob alternate sending bits according to the following high level strategy: 


\begin{itemize}
    \item A party (the $\speaker$) sends their next bit of $\pi_0$ \emph{until they are sure} that the other party (the $\listener$) has received their message. The $\listener$ gives feedback to the $\speaker$ by sending $0$ whenever the $\speaker$'s message was erased and $1$ whenever they receive it.
    \item The $\speaker$ signals the parties to switch roles: the $\speaker$ becomes the $\listener$ and the $\listener$ becomes the $\speaker$.
\end{itemize}

Because the adversary can only erase $\frac12-\epsilon$ of the total messages, there must be many pairs of rounds where \emph{both parties hear each other}. In each of these pairs of rounds, the $\speaker$ receives confirmation (in the form of a $1$) that the $\listener$ received the most recent message they sent. Then the two parties unerroneously make progress in their simulation of $\pi_0$.

Intuitively, the $\speaker$ has \emph{three} things to communicate: the bit $0$, the bit $1$, and switching roles. The challenge is to do this by sending only two bits. 

We can let the bits $0$ and $1$ mean \emph{more than two things} if we let the two bits mean different things when heard at different parts of the protocol. More specifically, we partition the protocol into $\approx \frac{n}{\epsilon}$ \emph{blocks} of $\approx \frac1\epsilon$ rounds. In each block, the first bit that the $\listener$ receives ($1$ or $0$) tells them whether the $\speaker$ is trying to send a bit or switch roles. In the case that the $\speaker$ is trying to send a bit, the bits received during the rest of the block tell the $\listener$ what the $\speaker$'s bit $b$ was.

\begin{itemize}
    \item In each block, if the $\speaker$ wants to send a bit $b$ to add to the transcript, they send $1$'s until they receive signal that the listener has heard it, at which point they send $b$ for the rest of the block.
    \item When they hear that the $\listener$ received the message $b$, they change strategies in the next block to signal the role switch: they send only $0$'s. For distinction, this version of $\speaker$ mode is called the $\passer$ mode.
    \item When the $\listener$ hears this $0$, i.e. the first message they receive in a block is a $0$, they know that they've received the bit $b$. To figure out what $b$ was, they look back at the last block where the first message they heard was $1$, and let $b=0$ if it was followed by any $0$'s and $b=1$ otherwise. Take note of a subtlety here -- the $\listener$ could have received more than one $1$ before the bit $b$, but this doesn't actually matter as they receive any $0$'s in the block if and only if $b=0$. 
    \item The $\passer$ (ex-$\speaker$) stays in $\passer$ mode, sending only $0$'s, until they hear a $1$ from the other party. This received $1$ either confirms the receipt of the role switch signal (a $0$ heard before any $1$'s in a block), or the party already received the switch signal earlier and is already in $\speaker$ mode; either way, the $\passer$ is free to switch to $\listener$ mode in the next block since the other party will be in $\speaker$ mode.
\end{itemize}

In this way, the roles of the $\speaker$ and $\listener$ have now switched and the protocol is in an entirely ``reset'' state with one more bit of the noiseless transcript $\pi_0$ having been conveyed.

To understand why this protocol is resilient to $\frac12-\epsilon$ erasures, we use the following fact: There are at least $\approx n$ blocks with $<\frac12-\epsilon$ erasures, and in each such block, there are at least a constant, say 2, pairs of consecutive messages that are unerased. In each such pair of messages, the protocol makes \emph{progress} via one party hearing the other and successfully confirming receipt of the message. (Formally, these two pairs of unerased consecutive messages are enough to ensure that at least one party completes the goal for that mode and advances modes at the end of the block.) 

For a formal description and analysis of this protocol, we refer the reader to Section~\ref{sec:erasure}.
\section{Preliminaries}

All definitions presented in this section are for the binary alphabet $\{ 0, 1 \}$. 

\paragraph{Notation.} In this work, we use the following notations.
\begin{itemize}
    \item The function $\Delta(x, y)$ represents the Hamming distance between $x$ and $y$.
    \item $x[i]$ denotes the $i$'th bit of a string $x \in \{ 0, 1 \}^*$.
    \item $x||y$ denotes the string $x$ concatenated with the string $y$.
    \item The symbol $\perp$ in a message represents the erasure symbol that a party might receive in the erasure model.
\end{itemize}

\subsection{Error Correcting Codes}

\begin{definition} [Error Correcting Code]
    A family of error correcting codes ($\ECC$) is a family of maps $\ECC = \{ \ECC_n : \{ 0, 1 \}^n \rightarrow \{ 0, 1 \}^{p(n)} \}_{n \in \bbN}$. An $\ECC$ has \emph{relative distance $\alpha > 0$} if for all $n \in \bbN$ and any $x \not= y \in \{ 0, 1 \}^n$,
    \[
        \Delta \left( \ECC_n(x), \ECC_n(y) \right) \ge \alpha p.
    \]
\end{definition}

Binary error correcting codes with relative distance $\approx \frac12$ are well known to exist with linear blowup in communication complexity, such that they can be encoded and decoded efficiently.

\begin{theorem} [\cite{GuruswamiS00}] \label{thm:ECC}
    For all $\epsilon > 0$, there exists an explicit family of error correcting codes $\ECC_\epsilon = \{ \ECC_{\epsilon, n} : \{ 0, 1 \}^n \rightarrow \{ 0, 1 \}^{p} \}_{n \in \bbN}$ with relative distance $\frac12 - \epsilon$ and maximum distance between any two elements $\frac12$\footnote{While the relevant theorem in the paper by Guruswami and Sudan does not directly state that the maximum distance between any two elements is $\frac12$, this follows from the fact that their construction is a concatenated code with the inner code as a Hadamard code (for which any two codewords have distance exactly $\frac12$).}, and with $p = p(n) = O\left( \frac{n}{\epsilon^3} \right)$. Moreover, there are explicit encoding and decoding algorithms that take time $C(\epsilon)n^{O(1)}$ for some constant $C(\epsilon)$. Correctness of decoding holds for up to $\frac14-\epsilon$ errors.
\end{theorem}


\subsection{Noise Resilient Interactive Communication} \label{sec:ip-def}

We formally define a non-adaptive interactive protocol along with error resilience to corruption. The two types of corruptions we will be interested in are erasures and bit flips. 

\begin{definition} [Non-Adaptive Interactive Coding Scheme] \label{def:ip}
    A two-party non-adaptive interactive coding scheme $\pi$ for a function $f(x, y) : \{ 0, 1 \}^n \times \{ 0, 1 \}^n \rightarrow \{ 0, 1 \}^o$ is an interactive protocol consisting of a fixed number of rounds, denoted $|\pi|$. In each round, a single party fixed beforehand sends a single bit to the other party. At the end of the protocol, each party outputs a guess $\in \{ 0, 1 \}^o$.
    
    
    We say that $\pi$ is resilient to $\alpha$ fraction of adversarial bit flips (resp. erasures) with probability $p$ if the following holds. For all $x,y \in \{ 0, 1 \}^n$, and for all online adversarial attacks consisting of flipping (resp. erasing) at most $\alpha \cdot |\pi|$ of the total rounds, with probability $\ge p$ Alice and Bob both output $f(x,y)$ at the end of the protocol. 
\end{definition}

\section{Binary Interactive Protocol Resilient to $\frac16$ Bit Flips} \label{sec:toggle}

In this section, we present a non-adaptive interactive protocol where Alice and Bob exchange inputs in a way that is resilient to $\frac16-\epsilon$ bit flips for any $\epsilon>0$. This result is optimal; no protocol is resilient to $\geq \frac16$ bit flips for all possible functions Alice and Bob might want to compute.

\begin{theorem} [\cite{EfremenkoGH16}]
    For large enough $n$, there exists a function $f(x, y)$ of Alice and Bob's inputs $x,y\in\{0,1\}^n$, such that any non-adaptive interactive protocol over the binary bit flip channel that computes $f(x, y)$ succeeds with probability at most $\frac12$ if a $\frac16$ fraction of the transmissions are corrupted.
\end{theorem}

As explained in Section~\ref{sec:overview}, in our protocol Alice and Bob each keep track of a string $\hat{y}$ and $\hat{x}$ respectively containing a guess for the other party's input, and a counter $c_A$ and $c_B$ respectively containing their confidence for the current guess. When a party receives a message, they decode it to obtain two parts: the other party's guess and instructions for how to update their own guess. They perform the corresponding updates to their own guess and then send their new guess along with instructions to the other party for how to update their guess. In order to formally state our protocol, we list a few definitions.

\subsection{Preliminaries and Definitions}

Throughout the protocol, Alice and Bob will send each other instructions, usually denoted $\delta$. To this end, we define two functions for how to use and create the instruction $\delta$.

\begin{definition}[$(z,c) \oplus \delta$]
We define $(z,c) \oplus \delta $ for $z \in \{ 0, 1 \}^{\leq n}$, $c\in \bbZ_{\geq 0}$ and $\delta\in \{0, 1, \rewind, \bullet\}$ as the update to $(z,c)$ induced by $\delta$. More specifically, we modify $z$ by the operation $\delta$ if $\delta \in \{0, 1, \rewind\}$ and increment the counter $c$ if $\delta=\bullet$. That is,
\begin{itemize}
    \item If $c = 0$ and $\delta \in \{ 0, 1 \}$, $|z| < n$, then $(z, c) \oplus \delta := (z||\delta, c)$.
    \item If $c = 0$ and $\delta = \rewind$, $|z| > 0$, then $(z, c) \oplus \leftarrow\ := (z[1:|z|-1], c)$. 
    \item If $c > 0$ and $\delta \in \{ 0, 1, \rewind \}$, then $(z, c) \oplus \leftarrow\ := (z, c-1)$.
    \item If $\delta = \bullet$, then $(z, c) \oplus \bullet := (z, c+1)$. 
    \item Otherwise, $(z, c) \oplus \delta := (z, c)$. 
\end{itemize}
\end{definition}

\begin{definition}[$\op_z$]
We define $\op_z(z')$ for $z \in \{ 0, 1 \}^{\le n}$ to be the instruction that brings $z'$ one bit closer to $z$ (or $\bullet$ if $z' = z$). That is, 
\begin{itemize}
    \item If $z'$ is a strict prefix of $z$, then $\op_z(z') := z[|z'| + 1]$.
    \item If $z'$ is not a prefix of $z$, then $\op_z(z') := \rewind$.
    \item If $z' = z$, then $\op_z(z') := \bullet$.
\end{itemize}
\end{definition}

Note that $(z', c) \oplus \op_z(z')$ either increases $c$ by $1$ if $z = z'$, and otherwise either decreases $c$ by $1$ if $c > 0$ or changes $z'$ to be one character closer to $z$.

Next, we define the error correcting code family $\ECC$ that Alice and Bob use in the protocol, and show that the desired $\ECC$ with the listed properties exists. For shorthand, we will denote the domain of the $\ECC$ as $$\Sigma=\{ 0, 1 \}^{\le n} \times \{ 0, 1 \}^{\le n} \times \{0, 1, \rewind, \bullet\}$$ throughout the section.

\begin{definition}[$\ECC$] \label{def:ecc}
For a given $\epsilon > 0$, we define the error correcting code family $$\ECC_\epsilon= \{ \ECC_{\epsilon,n} : \Sigma \rightarrow \{ 0, 1 \}^{M(n)} \}_{n \in \bbN}$$ with the following properties:
\begin{itemize}
    \item $M=O_\epsilon(n)$.
    \item For any $n \in \bbN$ and for any $(z_0, z'_0) \not= (z_1, z'_1) \in \{ 0, 1 \}^{\le n} \times \{ 0, 1 \}^{\le n}$ and $\delta_0, \delta_1 \in \{ 0, 1, \rewind, \bullet \}$, 
    \[
        \Delta \big( \ECC_{\epsilon,n}(z_0, z'_0, \delta_0), \ECC(z_1, z'_1, \delta_1) \big) \ge \left( \frac12 - \epsilon \right) \cdot M, \numberthis \label{eqn:ecc-1/2}
    \]
    \item For any $n \in \bbN$ and for any $z,z' \in \{ 0, 1 \}^{\le n} \times \{ 0, 1 \}^{\le n}$ and $\delta_0 \not= \delta_1 \in \{ 0, 1, \rewind, \bullet \}$, 
    \[
        \Delta \big( \ECC_{\epsilon,n}(z, z', \delta_0), \ECC(z, z', \delta_1) \big) \geq \frac23M. \numberthis \label{eqn:ecc-2/3}
    \]
    \item Decoding up to $\frac16 - \epsilon$ errors can be done in time $C(\epsilon)n^{O(1)}$ for some constant $C(\epsilon)$.
\end{itemize}
\end{definition}

\begin{claim} \label{claim:ecc}
For all $\epsilon>0$, an explicit error correcting code family $\ECC_\epsilon$ from Definition~\ref{def:ecc} exists. In other words, there exists an explicit error correcting code family that simultaneously satisfies all the properties listed.
\end{claim}

\begin{proof}
    For a fixed $n\in\bbN$, let $\ECC' : \{ 0, 1 \}^{\le n} \times \{ 0, 1 \}^{\le n} \rightarrow \{ 0, 1 \}^{M/3}$ be an efficiently encodable and decodable error correcting code with relative distance in the range $\left[ \frac12 - \epsilon, \frac12 \right]$ between any pair of codewords, where correctness of decoding holds for $\le \frac14 - \epsilon$ errors. This exists for some $M = O_\epsilon(n)$ by Theorem~\ref{thm:ECC}. Let $\DEC'$ be the corresponding decoding algorithm. We define
    \begin{align*}
        \ECC_{\epsilon,n}(z, z', \delta') = \begin{cases}
            \overline{\ECC'(z, z')} || \ECC'(z, z') || \ECC'(z, z'), & \delta' = 0 \\
            \ECC'(z, z') || \overline{\ECC'(z, z')} || \ECC'(z, z'), & \delta' = 1 \\
            \ECC'(z, z') || \ECC'(z, z') || \overline{\ECC'(z, z')}, & \delta' = \rewind \\
            \overline{\ECC'(z, z')} || \overline{\ECC'(z, z')} || \overline{\ECC'(z, z')}, & \delta' = \bullet,
        \end{cases}
    \end{align*}
    where $\bar{s}$ denotes the bitwise not of string $s$. 
    Then Equation~\eqref{eqn:ecc-1/2} holds because for any $(z_0, z'_0) \not= (z_1, z'_1)$, the relative distance between $\ECC'(z_0, z'_0)$ or $\overline{\ECC'(z_0, z'_0)}$ to $\ECC'(z_1, z'_1)$ or $\overline{\ECC'(z_1, z'_1)}$ is $\ge \frac12 - \epsilon$, and Equation~\eqref{eqn:ecc-2/3} holds because for fixed $z, z'$ the four codewords for $\delta' = 0, 1, \rewind, \bullet$ are distance $\frac23$ apart.

    It remains to show if a codeword $\ECC_{\epsilon}(z, z', \delta')$ is corrupted on $\le \left( \frac16 - \epsilon \right)$ locations to a string $s \in \{ 0, 1 \}^M$, then it can be correctly and efficiently decoded. Our decoding algorithm $\DEC(s)$ is as follows:
    \begin{quote}
        For a string $s=s_1s_2s_3$ where $s_i$ are each length $\frac{M}3$, consider each of the following strings:
        \begin{align*}
            S_0&=\overline{s_1}||s_2||s_3 \\
            S_1&=s_1||\overline{s_2}||s_3 \\
            S_{\rewind}&=s_1||s_2||\overline{s_3} \\
            S_{\bullet}&=\overline{s_1}||\overline{s_2}||\overline{s_3}.
        \end{align*} 
        For each $\delta \in \{ 0, 1, \rewind, \bullet \}$, denote $S_\delta = s^\delta_1||s^\delta_2||s^\delta_3$ and consider the string 
        \[
            S'_\delta=\maj(s^\delta_1[1],s^\delta_2[1],s^\delta_3[1])||\ldots || \maj(s^\delta_1[M/3],s^\delta_2[M/3],s^\delta_3[M/3]) \in \{ 0, 1 \}^{M/3}.
        \]
        Compute $(z_\delta, z'_\delta) \gets \DEC'(S'_\delta)$, and let $d_\delta = \frac{1}{M} \Delta(\ECC(z_\delta, z'_\delta, \delta), s)$. If $d_\delta \le \left(\frac16 - \epsilon\right)$, output $(z_\delta, z'_\delta, \delta)$. 
        
        If $d_\delta > \left(\frac16 - \epsilon\right)$ for all $\delta \in \{ 0, 1, \rewind, \bullet \}$, then output $\emptyset$.
    \end{quote}
    
    We now show correctness of this decoding algorithm if there are fewer than $\frac16 - \epsilon$ corruptions. First note that for any string $s$, since any two codewords $\ECC(z_{\delta_0}, z'_{\delta_0}, \delta_0), \ECC(z_{\delta_1}, z'_{\delta_1}, \delta_1)$ with $\delta_0 \not= \delta_1$ are at least relative distance $\frac12 - \epsilon$ apart, $d_\delta$ can be $\le \frac16 - \epsilon$ for at most one value of $\delta$. This means that the output condition is satisfied for at most one codeword of the four.
    
    If $s$ is relative distance $\le \frac16 - \epsilon$ to a codeword $\ECC_\epsilon(z, z', \delta')$, then $S_{\delta'}$ is relative distance $\le \frac16 - \epsilon$ to $\ECC'(z, z') || \ECC'(z, z') || \ECC'(z, z')$. This means that $S'_{\delta'}$ and $\ECC(z, z')$ differ on at most $\frac12 \cdot ( \frac16 - \epsilon ) \cdot M$ locations, which is $< \frac14 - \epsilon$ fraction of the $\frac{M}{3}$ locations, since the adversary must corrupt two out of three values $s^{\delta'}_1[j], s^{\delta'}_2[j], s^{\delta'}_3[j]$ in order for $S'_{\delta'}[j] \not= \ECC(z, z')[j]$. Thus, $(z, z') = \DEC'(S'_{\delta'})$, and the decoding algorithm outputs $(z, z', \delta')$.

    
\end{proof}

For the rest of the section, $\ECC$ will denote the error correcting code $\ECC_{\epsilon,n}$ as defined in Definition~\ref{def:ecc} when $n$ and $\epsilon$ are clear.

\subsection{Formal Protocol}
We are now ready to formally state the protocol.

\protocol{Randomized Protocol Resilient to $\frac16 - 2\epsilon$ Corruptions}{1/6}{
    Fix $n \in \bbN, \epsilon>0$. Suppose Alice and Bob's private inputs are $x, y \in \{ 0, 1 \}^n$ respectively, such that $y[1] = 0$. The protocol consists of $T = \frac{n}{\epsilon}$ (assume $T$ is even) messages numbered $1, \dots, T$, each of length $M$. Alice sends the odd messages and Bob sends the even. For our convenience, Alice and Bob both agree that the $0$'th message is $\ECC(\emptyset,\emptyset,0)$, sent uncorrupted by Bob to Alice.
    
    \vspace{0.075in}
    
    Alice and Bob track a private guess for the other party's input, denoted $\hat{y} \in \{ 0, 1 \}^{\le n}$ and $\hat{x} \in \{ 0, 1 \}^{\le n}$ respectively, both initialized at $\emptyset$ at the beginning of the protocol. They also each have a personal counter $c_A$ and $c_B$ respectively, containing their confidence in their current guess $\hat{y}$ or $\hat{x}$ respectively, initialized at $0$. 
    
    \vspace{0.075in}
    
    In what follows, we describe Alice's behavior. Bob's behavior is identical, except replacing $\hat{y}, c_A$ by $\hat{x}, c_B$ and notationally switching $x$ and $y$ in general (notably sending $\ECC(\hat{x},y^*,\delta^*)$ instead of $\ECC(\hat{y},x^*,\delta^*)$). At the end of the protocol, Alice and Bob will output $\hat{y}$ and $\hat{x}$ respectively.
    
    \begin{center}
    {\bf \fbox{Alice}}
    \end{center}

        Alice has just received a message $m$ from Bob. She first tries to set $(x^*, y^*, \delta^*) \in \Sigma$ and $p^* \in [0, 1]$ as follows. If no condition is satisfied, we say that she does not set $(x^*, y^*, \delta^*)$. In what follows, we let $d_m(x', y', \delta')$ denote $\frac1M \cdot \Delta(m, \ECC(x', y', \delta'))$.
        \begin{enumerate}
        \item \label{item:one}
           She sets $(x^*, y^*, \delta^*) \gets (x, \hat{y}, \delta)$ if $d_m(x, \hat{y}, \delta) < \frac13$ by checking all $\delta \in \{ 0, 1, \rewind, \bullet \}$ individually.
            If she sets $(x^*, y^*, \delta^*)$ in this way, she also sets 
            \begin{align*}
                p^* = 1 - 3 d_m(x, \hat{y}, \delta).
            \end{align*}
            Note that $d_m(x, \hat{y}, \delta) < \frac13$ for at most one $\delta$, since the codewords $\ECC(x, \hat{y}, \delta)$ all have relative distance $\frac23$.
        \item 
            Otherwise, if $\DEC(m) \not= \emptyset$, she sets $(x^*, y^*, \delta^*) \gets \DEC(m)$. She sets 
            \[
                p^* = 0.5 - 3 d_m(x^*, y^*, \delta^*)
            \]
            Recall that $\DEC(m) \not= \emptyset$ only if $d_m(x^*, y^*, \delta^*) \le \frac16 - \epsilon$. 
        \end{enumerate}
        
        Next, based on whether or not Alice set $(x^*, y^*, \delta^*)$, she does the following:
        
        \begin{itemize}
        \item 
            If she set $(x^*, y^*, \delta^*)$, she does the following update with probability $p^*$ (doing nothing if no condition is satisfied):
            
            \begin{itemize}
            \item 
                If $y^* = \hat{y}$, then $(\hat{y},c_A) \gets (\hat{y},c_A) \oplus \delta^*$.
            \item 
                If $y^* \not= \hat{y}$ and $\delta^*=\bullet$, then $(\hat{y},c_A) \gets (\hat{y},c_A) \oplus \op_{y^*}(\hat{y})$.
            \end{itemize}
            
            She sends $\ECC(\hat{y},x^*,\op_x(x^*)$.
            
        \item 
            Otherwise, if she did not sent $(x^*, y^*, \delta^*)$, she does nothing to $\hat{y}, c_A$ and sends $\ECC(\hat{y}, x, \bullet)$.
        \end{itemize}
        
        
}

\subsection{Analysis}

We now state our main error resilience theorem.

\begin{theorem} \label{thm:1/6}
    Protocol~\ref{prot:1/6} is resilient to a $\left(\frac16 - 2\epsilon\right)$-fraction of errors with probability $1 - 2\exp \left( \frac{-\epsilon n}{100} \right)$.\footnote{We only consider $\epsilon$ sufficiently small (say $<0.1$).}
\end{theorem}

To prove Theorem~\ref{thm:1/6}, we analyze the effects of corruption on the \emph{good} and \emph{bad updates} Alice/Bob make. We begin by defining good and bad updates. After receiving a message from Bob, Alice updates (with some probability) her values of $c_A$ and $\hat{y}$ (if these values both stay the same, we say that Alice did not make an update). We say that an update is \emph{good} if her new value of $(\hat{y},c_A)$ is $(\hat{y},c_A) \oplus \op_y(\hat{y})$. An update is otherwise \emph{bad}. We similarly define good and bad updates for Bob.

\noindent For each $t \in [0, \dots, T]$, we define the following potential functions:
\begin{itemize}
    \item $\psi^A_t$ is defined to be the total number of good updates minus the number of bad updates Alice has done in response to messages $0, \dots, t$.
    \item $\psi^B_t$ is defined to be the total number of good updates minus the number of bad updates Bob has done in response to messages $0, \dots, t$.
\end{itemize}
For instance, since there's an agreed-upon $0$'th message from Bob to Alice, $\bbE[\psi^A_0] = 0.5$ and $\psi^B_0 = 0$. For $t = -1$, we let $\psi^A_{-1} = \psi^B_{-1} = 0$.

\begin{lemma} \label{lemma:N}
    After message $t$, $\psi_t^A \ge n$ if and only if Alice's value of $\hat{y}$ is equal to $y$, and $\psi^B_t \ge n$ if and only if Bob's value of $\hat{x}$ is equal to $x$.
    
    In particular, at the end of the protocol, Alice outputs $\hat{y} = y$ if and only if $\psi^A_T \ge n$, and Bob outputs $\hat{x}$ if and only if $\psi^B_T \ge n$.
\end{lemma}

\begin{proof}
    We prove this for Alice, as the proof for Bob is identical. We keep track of the number of good updates Alice must do to have $\hat{y} = y$. At the beginning of the protocol, since $\hat{y} = \emptyset$, Alice needs to perform $n$ good updates (appending the $n$ bits of $y$) so that $\hat{y} = y$. We show that any good update decreases this number by $1$, and any bad update increases this number by $1$. Then, at any point, $\hat{y} = y$ if and only if the number of good updates minus the number of bad updates is at least $n$.
    
    To show that every bad update increases this number by 1, we show that a bad update (i.e., any update other than $(\hat{y},c_A) \gets (\hat{y},c_A) \oplus \op_y(\hat{y})$), when followed by the good update $(\hat{y},c_A) \gets (\hat{y},c_A) \oplus \op_y(\hat{y})$, results back in the original value of $(\hat{y},c_A)$. If the bad update appends a bit to $\hat{y}$, then the new value of $\hat{y}$ must not be a prefix of $y$. Then the good update $\rewind$ undoes this. If the bad update deletes the last bit of $\hat{y}$ incorrectly, then re-appending this bit undoes this. If the bad update increases $c_A$ incorrectly, then $\hat{y} \not= y$, so the next good update is $\op_y(\hat{y}) \not= \bullet$ which causes $c_A$ to decrease by $1$. If the bad update decreases $c_A$ incorrectly, then $\hat{y} = y$, and the next good update is $\op_y(\hat{y}) = \bullet$ which increases $c_A$ by $1$.
    
    
\end{proof}

For Alice, we group each consecutive pair of Alice-to-Bob and Bob-to-Alice messages (i.e., messages $2k-1$ and $2k$). For Bob, we group each consecutive pair of Bob-to-Alice and Alice-to-Bob messages (i.e., messages $2k-2$ and $2k-1$), starting with the unsent message $0$ which we recall is understood to be $\ECC(\emptyset,\emptyset,0)$. We define the following potential function for each $k \in [0, T/2]$: 
\begin{itemize}
    \item $\Psi^A_k = \psi^A_{2k} + \min \{ \psi^B_{2k} + \min \{ \rho^B_{2k+1}, 0.5 \}, n \}$
    \item $\Psi^B_k = \psi^B_{2k-1} + \min \{ \psi^A_{2k-1} + \min \{ \rho^A_{2k}, 0.5 \}, n \}$,
\end{itemize}
where $\rho^A_t$ and $\rho^B_t$ are defined as follows:
\begin{itemize}
    \item For even $t$, $\rho^A_t$ is the expected number of good updates minus the number of bad updates that Alice will do in response to message $t$, given messages $1,\dots,t-1$, if message $t$ from Bob is uncorrupted.
    \item For odd $t$, $\rho^B_t$ is the expected number of good updates minus the number of bad updates that Bob will do in response to message $t$, given messages $1,\dots,t-1$, if message $t$ from Alice is uncorrupted. For instance, since Alice sends $\ECC(\hat{y}, \emptyset, x[1])$ as her first message, $\bbE[\rho^B_1] = 0.5$.
\end{itemize} 

\begin{lemma} \label{lemma:0.5-3alpha}
    If an $\alpha_{2k-1}$ fraction of message $2k-1$ and an $\alpha_{2k}$ fraction of message $2k$ is corrupted, then 
    \[
        \bbE [\Psi^A_k - \Psi^A_{k-1}] \ge 1 - 6\epsilon - 3\alpha_{2k-1} - 3\alpha_{2k}. \numberthis \label{eqn:alice-0.5-3alpha}
    \]
    If an $\alpha_{2k-2}$ fraction of message $2k-2$ and an $\alpha_{2k-1}$ fraction of message $2k-1$ is corrupted, then
    \[
        \bbE [\Psi^B_k - \Psi^B_{k-1}] \ge 1 - 6\epsilon - 3\alpha_{2k-2} - 3\alpha_{2k-1}. \numberthis \label{eqn:bob-0.5-3alpha}
    \]
\end{lemma}
    
We delay the proof of Lemma~\ref{lemma:0.5-3alpha} to Section~\ref{sec:casework} as it is quite long and instead use it to complete the proof of Theorem~\ref{thm:1/6}.

\begin{proof}[Proof of Theorem~\ref{thm:1/6}]
    Consider any adversarial corruption of Protocol~\ref{prot:1/6} consisting of fewer than $\frac16 - 2\epsilon$ corruptions. We will show that Alice and Bob must both output the other party's input correctly. Let $\alpha_1, \dots, \alpha_{T}$ denote the fractional number of corruptions in messages $1, \dots, T$, such that $\alpha_1 + \dots + \alpha_{T} < \left( \frac16 - 2\epsilon \right) \cdot T$. By default, we let $\alpha_0 = 0$. For $k \in [T/2]$, we define the random variables
    \begin{align*}
        \Phi^A_k &= \Psi^A_k - k + 6k\epsilon + \sum_{i=0}^{2k} 3\alpha_i, \\
        \Phi^B_k &= \Psi^B_k - k + 6k\epsilon + \sum_{i=0}^{2k-1} 3\alpha_i.
    \end{align*}
    Then, $\Phi^A_0 = \Psi^A_0 \in \{ 0.5, 1.5 \}$ and $\Phi^B_0 = \Psi^B_0 = 0.5$.
    
    By Lemma~\ref{lemma:0.5-3alpha}, 
    \begin{align*}
        \bbE[\Phi^A_k] &= \bbE \left[ \Psi^A_k - k + 6k\epsilon + \sum_{i=0}^{2k} 3\alpha_i \right] \\
        &\ge \bbE \left[ \Psi^A_{k-1} - (k-1) +6(k-1)\epsilon + \sum_{i=0}^{2k-2} 3\alpha_i \right] \\
        &= \bbE[\Phi^A_{k-1}],
    \end{align*}
    so $\Phi^A_k$ is a submartingale with bounded distance 
    \begin{align*}
        |\Phi^A_k - \Phi^A_{k-1}| 
        &= \left| \Psi^A_k - \Psi^A_{k-1} - 1 +6\epsilon + 3\alpha_{2k-1} + 3\alpha_{2k} \right| \\
        &\le \left| U^A_{2k} \right| + \left| U^B_{2k-1} \right| + \left| \rho^B_{2k+1} \right| + \left| \rho^B_{2k-1} \right| + \left|- 1 +6\epsilon + 3\alpha_{2k-1} + 3\alpha_{2k} \right| \\
        &< 10,
    \end{align*}
    where $U^A_{2k}$ is $+1$, $-1$, or $0$ if Alice made a good, bad, or no update respectively in response to message $2k$, and $U^B_{2k}$ is defined the same way but for Bob in response to message $2k-1$. 
    Similarly, $\Phi^B_k$ is a submartingale with bounded distance $< 10$.
    
    Note that if $\Psi^A_{T/2} \ge 2n$, then $\psi^A_T = \Psi^A_{T/2} - \min \{ \psi^B_T, n \} \ge n$, meaning by Lemma~\ref{lemma:N} that Alice holds $\hat{y} = y$ at the end of the protocol. Thus, by Azuma's inequality, the probability that Alice outputs correctly at the end of the protocol is at least
    \begin{align*}
        \Pr \left[ \Psi^A_{T/2} \ge 2n \right]
        &= 1 - \Pr \left[ \Phi^A_{T/2} - \Phi^A_0 < 2n - \frac{T}{2} + 3T\epsilon + \sum_{i=0}^{T} 3\alpha_i - \Phi^A_0 \right] \\
        &\ge 1 - \Pr \left[ \Phi^A_{T/2} - \Phi^A_0 < 2n - \frac{T}{2} + 3T\epsilon + 3 \cdot \left( \frac16 - 2\epsilon \right) \cdot T \right] \\
        &\ge 1 - \Pr \left[ \Phi^A_{T/2} - \Phi^A_0 < -n \right] \\
        &\ge 1 - \exp \left( \frac{-\epsilon n}{100} \right).
    \end{align*}
    The same calculation for Bob shows that Bob outputs correctly at the end of the protocol with probability at least $1 - \exp \left( \frac{-\epsilon n}{100} \right)$ as well. Then by a union bound, the probability that both parties output correctly is at least
    \[
        1 - 2 \cdot \exp \left( \frac{-\epsilon n}{100} \right).
    \]
    
\end{proof}

\subsubsection{Proof of Lemma~\ref{lemma:0.5-3alpha}}\label{sec:casework}

We only prove Inequality~\eqref{eqn:alice-0.5-3alpha}, as the proof for Inequality~\eqref{eqn:bob-0.5-3alpha} is identical. 
    
Define $\Psi_k$ to be $\psi^A_{2k} + \rho^A_{2k} + \min \{ \psi^B_{2k-1}, n \}$. We will first determine the value of $\bbE[ \Psi^A_k - \Psi_k ]$, which we will then use to complete the proof of the lemma.
    
\begin{claim} \label{claim:PsiA-Psi}
    If $\psi^B_{2k-1} < n$, then 
    \[
        \bbE [ \Psi^A_k - \Psi_k ] \ge 0.5 - 3\epsilon - 3\alpha_{2k}. \numberthis \label{eqn:PsiA-Psi--less-n}
    \]
    If $\psi^B_{2k-1} \ge n$, then
    \[
        \bbE [ \Psi^A - \Psi_k ] \ge -3\epsilon -3\alpha_{2k}. \numberthis \label{eqn:PsiA-Psi--greater-n}
    \]
\end{claim}

\begin{proof}

    Define the random variable $U^A_{2k}$ to be $1$ if Alice makes a good update, $-1$ if she makes a bad update, and $0$ if she makes no update in response to message $2k$. Let $\ECC(x_{2k},y_{2k},\delta_{2k})$ be Bob's intended message for message $2k$, and let $m_{2k}$ be the message Alice receives. Let $(x^*_{2k}, y^*_{2k}, \delta^*_{2k}) \in \Sigma$ with a corresponding $p^*_{2k}$ be what Alice computes in her protocol, if they exist. Moreover, let $d^*_{2k} = d_{m_{2k}}(x^*_{2k}, y^*_{2k}, \delta^*_{2k})$.
    In order to show Equations~\eqref{eqn:PsiA-Psi--less-n} and~\eqref{eqn:PsiA-Psi--greater-n}, we will show they hold for any value of $(x_{2k}, y_{2k}, \delta_{2k})$, which implies they hold in expectation.
    
    \paragraph{Proof of Equation~\eqref{eqn:PsiA-Psi--less-n}.} We first show that if $\psi^B_{2k-1} < n$, then $\bbE[\Psi^A_k - \Psi_k] \ge 0.5 - 3\epsilon - 3\alpha_{2k}$. 
    To start, we have
    \begin{align*}
        \bbE[\Psi^A_k - \Psi_k] 
        &= \bbE \left[ \psi^A_{2k} + \min \{ \psi^B_{2k} + \min \{ \rho^B_{2k+1}, 0.5 \}, n \} - \psi^A_{2k-1} - \rho^A_{2k} - \min \{ \psi^B_{2k-1}, n \} \right] \\
        &= \bbE \left[ \psi^A_{2k} + \psi^B_{2k} + \min \{ \rho^B_{2k+1}, 0.5 \} - \psi^A_{2k-1} - \rho^A_{2k} - \psi^B_{2k-1} \right] \\
        &= \bbE [U^A_{2k}-\rho^A_{2k}+\min \{ \rho^B_{2k+1}, 0.5 \}],
    \end{align*}
    as $\psi^B_{2k-1} = \psi^B_{2k} < n$, implying that $\psi^B_{2k} + \min\{ \rho^B_{2k+1}, 0.5 \}, \psi^B_{2k-1} \le n$.
    
    We split the analysis into cases. In each case, we bound the three values $\bbE [U^A_{2k}], \rho^A_{2k}, \rho^B_{2k+1}$ separately, and combine them to bound $\bbE[\Psi^A_k - \Psi_k]$. Since $\psi^B_{2k-1} < n$, by Lemma~\ref{lemma:N}, Bob's value of $\hat{x}$ after receiving message $2k-1$ is not $x$, i.e. $x_{2k} \not= x$. Thus, $\rho^A_{2k} \le 0.5$. Furthermore, $\rho^B_{2k+1} \ge 0$ because any uncorrupted message Alice sends cannot cause Bob to perform a bad update.
    If no other constraint on $\rho^A_{2k}$ or $\min \{ \rho^B_{2k+1}, 0.5 \}$ is used in the final calculation, we omit its computation.

    \begin{caseof}
    \begin{mdframed}[topline=false,rightline=false,bottomline=false]
    \case{$(x^*_{2k},y^*_{2k},\delta^*_{2k})$ does not exist.}{
    \vspace{0.075in}
    \begin{itemize}
        \item $\bbE[U^A_{2k}]=0$ because Alice does not update.
        \item $\rho^B_{2k+1}\geq 0.5$ because she replies with $\ECC(\hat{y}, x, \bullet)$, and $(\hat{x},c_B) \oplus \op_x(\hat{x})$ is a positive update.
        \item $\alpha_{2k}\geq \left(\frac16-\epsilon\right)$ or we would have $(x_{2k},y_{2k},\delta_{2k}) = (x^*_{2k},y^*_{2k},\delta^*_{2k})$.
    \end{itemize}
    \vspace*{-\baselineskip}
        \begin{align*}
            \implies\bbE[\Psi^A_k - \Psi_k] &= \bbE [U^A_{2k}-\rho^A_{2k}+\min \{ \rho^B_{2k+1}, 0.5 \}] \\
            & \geq 0-0.5+0.5=0 \\
            & \geq 0.5-3\epsilon-3\alpha_{2k}.
        \end{align*}
    }
    \case {$(x_{2k},y_{2k},\delta_{2k}) = (x^*_{2k},y^*_{2k},\delta^*_{2k})$ and ($y_{2k}=\hat{y}$ or $\delta_{2k}=\bullet$).}{ 
    \vspace{0.075in}
    \begin{itemize}
        \item $\bbE[U^A_{2k}] \ge 0.5 - 3\alpha_{2k}$. Alice makes a good update with probability $p^*_{2k} = 0.5 - 3d^*_{2k}>0$, and $d^*_{2k}\le \alpha_{2k}$, so $\bbE[U^A_{2k}] = 0.5 - 3d^*_{2k} \ge 0.5 - 3\alpha_{2k}$.
        \item $\rho^B_{2k+1}\geq 0.5$ because she replies with $\ECC(\hat{y}, x_{2k}, \op_x(x_{2k}))$, and $(\hat{x}=x_{2k},c_B) \oplus \op_x(x_{2k})$ is a positive update since $x_{2k} = \hat{x}$.
    \end{itemize}
    \vspace*{-\baselineskip}
        \begin{align*}
            \implies\bbE[\Psi^A_k - \Psi_k] &= \bbE [U^A_{2k}-\rho^A_{2k}+\min \{ \rho^B_{2k+1}, 0.5 \}] \\
            &\geq (0.5-3\alpha_{2k})-0.5+0.5 \\
            &\geq 0.5-3\epsilon-3\alpha_{2k}.
        \end{align*}
    }
    \case {$(x_{2k},y_{2k},\delta_{2k}) = (x^*_{2k},y^*_{2k},\delta^*_{2k})$ and ($y_{2k}\neq \hat{y}$ and $\delta_{2k}\neq \bullet$).}
    { 
    \vspace{0.075in}
    \begin{itemize}
        \item $U^A_{2k}=0$ because Alice will never change $(\hat{y},c_A)$ in response to $(x^*_{2k},y^*_{2k},\delta^*_{2k})$ with the given constraints.
        \item $\rho^A_{2k}=0$ because again Alice will never change $(\hat{y},c_A)$ in response to $(x_{2k},y_{2k},\delta_{2k})$.
        \item $\rho^B_{2k+1}\geq 0.5$ because she replies with $\ECC(\hat{y}, x_{2k}, \op_x(x_{2k}))$.
    \end{itemize}
    \vspace*{-\baselineskip} 
        \begin{align*}
            \implies \bbE[\Psi^A_k - \Psi_k] &= \bbE [U^A_{2k}-\rho^A_{2k}+\min \{ \rho^B_{2k+1}, 0.5 \}] \\
            &\geq 0-0+0.5 \\
            &\geq 0.5-3\epsilon-3\alpha_{2k}.
        \end{align*}
    }
    \case {$(x_{2k},y_{2k},\delta_{2k}) \neq (x^*_{2k},y^*_{2k},\delta^*_{2k})$ and ($x^*_{2k}=x$ and $y^*_{2k} = \hat{y}$).}{ 
    \vspace{0.075in}
    \begin{itemize}
        \item $\bbE [U^A_{2k}] \geq 0.5-3\epsilon-3\alpha_{2k}$. Alice makes a bad update with probability at most $p^*_{2k} = 1-3d^*_{2k}>0$ (``at most'' because she may make a good update or no update). Substituting $d^*_{2k} \geq (\frac12-\epsilon)-\alpha_{2k}$ which follows from Equation~\eqref{eqn:ecc-1/2} gives the desired result.
        \item $\rho^B_{2k+1}\geq 0.5$ because she replies with $\ECC(\hat{y}, x, \bullet)$.
    \end{itemize}
    \vspace*{-\baselineskip}
        \begin{align*}
            \implies \bbE[\Psi^A_k - \Psi_k] &= \bbE [U^A_{2k}-\rho^A_{2k}+\min \{ \rho^B_{2k+1}, 0.5 \}] \\
            &\geq (0.5-3\epsilon-3\alpha_{2k}) - 0.5 + 0.5 \\
            &=0.5-3\epsilon-3\alpha_{2k}.
        \end{align*}
    }
    \case {$(x_{2k},y_{2k},\delta_{2k}) \neq (x^*_{2k},y^*_{2k},\delta^*_{2k})$ and ($x^*_{2k}\neq x$ or $y^*_{2k} \neq \hat{y}$).}{ 
    \vspace{0.075in}
    \begin{itemize}
        \item $\bbE[U^A_{2k}] \ge 1 - 3\epsilon - 3\alpha_{2k}$. Alice makes a bad update with probability at most $p^*_{2k} = 0.5-3d^*_{2k}$. Substituting $d^*_{2k} \geq \left(\frac12-\epsilon\right) - \alpha_{2k}$ gives the desired result.
    \end{itemize}
    \vspace*{-\baselineskip}
        \begin{align*}
            \implies\bbE[\Psi^A_k - \Psi_k] &= \bbE [U^A_{2k}-\rho^A_{2k}+\min \{ \rho^B_{2k+1}, 0.5 \}] \\
            &\geq (1-3\epsilon - 3\alpha_{2k}) - 0.5 + 0 \\
            &= 0.5 - 3\epsilon - 3\alpha_{2k}.
        \end{align*}
    }
    \end{mdframed}
    \end{caseof}
    
    \noindent
    \paragraph{Proof of Equation~\eqref{eqn:PsiA-Psi--greater-n}.} Next, we show that if $\psi^B_{2k-1} \ge n$, then $\bbE[ \Psi^A_k - \Psi_k ] \ge -3\alpha_{2k}$.
    Since $\psi^B_{2k-1} = \psi^B_{2k} \ge n$, we have that
    \begin{align*}
        \bbE[\Psi^A_k - \Psi_k] &= \bbE \left[ \psi^A_{2k} + \min \{ \psi^B_{2k} + \min \{ \rho^B_{2k+1}, 0.5 \}, n \} - \psi^A_{2k-1} - \rho^A_{2k} -  \min \{ \psi^B_{2k-1}, n \} \right] \\
        &= \bbE [U^A_{2k}-\rho^A_{2k}].
    \end{align*}
    Again, we split the analysis into cases. In each case we compute $\bbE [U^A_{2k}]$ and $\rho^A_{2k}$ separately. Since $\psi^B_{2k-1} \ge n$, by Lemma~\ref{lemma:N}, $x_{2k} = x$. We have that $\rho^A_{2k} \le 1$ (Alice cannot perform more than one good update in response to any possible message sent by Bob).
    
    \begin{caseof}
    \begin{mdframed}[topline=false,rightline=false,bottomline=false]
    \case {$(x^*_{2k},y^*_{2k},\delta^*_{2k})$ does not exist.}{ 
    \vspace{0.075in}
    \begin{itemize}
        \item $\bbE[U^A_{2k}]=0$ because Alice does not update.
        \item $\rho^A_{2k} \leq 3\epsilon + 3\alpha_{2k}$. If $y_{2k}=\hat{y}$, then $\rho^A_{2k} \leq 1$ and $\alpha_{2k}>\frac13$ since otherwise $(x^*_{2k},y^*_{2k},\delta^*_{2k})$ would've been $(x_{2k}, y_{2k}, \delta_{2k})$. 
        Else if $y_{2k} \not= \hat{y}$, $\rho^A_{2k} \leq 0.5$ and $\alpha_{2k}>\left(\frac16-\epsilon\right)$. Regardless, the result follows.
    \end{itemize}
    \vspace*{-\baselineskip}
        \begin{align*}
            \implies \bbE[\Psi^A_k - \Psi_k] &= \bbE [U^A_{2k}-\rho^A_{2k}] \\
            &\geq 0-3\epsilon-3\alpha_{2k} \\
            &\geq-3\epsilon-3\alpha_{2k}.
        \end{align*}
    }
    \case {$(x_{2k},y_{2k},\delta_{2k}) = (x^*_{2k},y^*_{2k},\delta^*_{2k})$ and $y_{2k}=\hat{y}$.}{ 
    \vspace{0.075in}
    \begin{itemize}
        \item $\bbE [U^A_{2k}]\geq 1- 3\alpha_{2k}$, because Alice makes a good update with probability $p^*_{2k} = 1 - 3d^*_{2k}$ and $\alpha_{2k}\geq d^*_{2k}$.
    \end{itemize}
    \vspace*{-\baselineskip} 
        \begin{align*}
            \implies \bbE[\Psi^A_k - \Psi_k] &= \bbE [U^A_{2k}-\rho^A_{2k}] \\
            &\geq 1-3\alpha_{2k}-1 \\
            &\geq-3\epsilon-3\alpha_{2k}.
        \end{align*}
    }
    \case {$(x_{2k},y_{2k},\delta_{2k}) = (x^*_{2k},y^*_{2k},\delta^*_{2k})$ and ($y_{2k} \not= \hat{y}$ and $\delta_{2k}=\bullet$).}{
    \vspace{0.075in}
    \begin{itemize}
        \item $\bbE [U^A_{2k}]\geq 0.5-3\alpha_{2k}$, because Alice makes a good update with probability $p^*_{2k} = 0.5 - 3d^*_{2k}$ and $\alpha_{2k}\geq d^*_{2k}$.
        \item $\rho^A_{2k} \le 0.5$ since $y_{2k} \not= \hat{y}$.
    \end{itemize}
    \vspace*{-\baselineskip} 
        \begin{align*}
            \implies \bbE[\Psi^A_k - \Psi_k] &= \bbE [U^A_{2k}-\rho^A_{2k}] \\
            &\geq 0.5-3\alpha_{2k}-0.5 \\
            &\geq -3\epsilon-3\alpha_{2k}.
        \end{align*}
    }
    \case {$(x_{2k},y_{2k},\delta_{2k}) = (x^*_{2k},y^*_{2k},\delta^*_{2k})$ and ($y_{2k}\neq \hat{y}$ and $\delta_{2k}\neq \bullet$).}{ 
    \vspace{0.075in}
    \begin{itemize}
        \item $U^A_{2k}=0$ because Alice will never change $(\hat{y},c_A)$ in response to $(x^*_{2k},y^*_{2k},\delta^*_{2k})$.
        \item $\rho^A_{2k}=0$ for the same reason.
    \end{itemize}
    \vspace*{-\baselineskip}
    \begin{align*}
        \implies \bbE[\Psi^A_k - \Psi_k] &= \bbE [U^A_{2k}-\rho^A_{2k}] \\
        &\geq 0-0 \\
        &\geq -3\epsilon -3\alpha_{2k}.
    \end{align*}
    }
    \case {$(x_{2k},y_{2k},\delta_{2k}) \neq (x^*_{2k},y^*_{2k},\delta^*_{2k})$ and ($x^*_{2k}=x$ and $y^*_{2k} = \hat{y} = y_{2k}$).}{ 
    \vspace{0.075in}
    \begin{itemize}
        \item $\bbE [U^A_{2k}]\geq 1 -3\alpha_{2k}$. Alice makes a bad update with probability $p^*_{2k} = 1-3d^*_{2k}>0$. Substituting $d^*_{2k} \ge \frac23 - \alpha_{2k}$ gives the desired result.
    \end{itemize}
    \vspace*{-\baselineskip}
    \begin{align*}
        \implies\bbE[\Psi^A_k - \Psi_k] &= \bbE [U^A_{2k}-\rho^A_{2k}] \\
        &\geq (1-3\alpha_{2k}) - 1 \\
        &\geq -3\epsilon-3\alpha_{2k}.
    \end{align*}
    }
    \case {$(x_{2k},y_{2k},\delta_{2k}) \neq (x^*_{2k},y^*_{2k},\delta^*_{2k})$ and ($x^*_{2k}=x$ and $y^*_{2k} = \hat{y} \neq y_{2k}$).}{ 
    \vspace{0.075in}
    \begin{itemize}
        \item $\bbE [U^A_{2k}] \geq 0.5-3\epsilon - 3\alpha_{2k}$. Alice makes a bad update with probability at most $\le p^*_{2k} = 1-3d^*_{2k}>0$. Substituting $d^*_{2k} \ge (\frac12 -\epsilon)- \alpha_{2k}$ gives the desired result.
        \item $\rho^A_{2k} \le 0.5$ since $y_{2k} \not= \hat{y}$.
    \end{itemize}
    \vspace*{-\baselineskip} 
    \begin{align*}
        \implies \bbE[\Psi^A_k - \Psi_k] &= \bbE [U^A_{2k}-\rho^A_{2k}] \\
        &\geq (0.5-3\epsilon - 3\alpha_{2k}) - 0.5 \\
        &= -3\epsilon -3\alpha_{2k}.
    \end{align*}
    }
    \case {$(x_{2k},y_{2k},\delta_{2k}) \neq (x^*_{2k},y^*_{2k},\delta^*_{2k})$ and ($x^*_{2k}\not= x$ or $y^*_{2k} \not= \hat{y}$).}{ 
    \vspace{0.075in}
    \begin{itemize}
        \item $\bbE [U^A_{2k}] \geq 1-3\epsilon-3\alpha_{2k}$. Alice makes a bad update with probability at most $p^*_{2k} = \frac12-3d^*_{2k}>0$. Substituting $d^*_{2k} \ge (\frac12-\epsilon) - \alpha_{2k}$ gives the desired result.
    \end{itemize}
    \vspace*{-\baselineskip}
    \begin{align*}
        \implies \bbE[\Psi^A_k - \Psi_k] &= \bbE [U^A_{2k}-\rho^A_{2k}] \\
        &\geq (1-3\epsilon-3\alpha_{2k}) - 1 \\
        &=-3\epsilon-3\alpha_{2k}.
    \end{align*}
    }
    
    \end{mdframed}
    \end{caseof}

\end{proof}

\noindent
We return now to the proof of Lemma~\ref{lemma:0.5-3alpha}. Recall that we want to show 
\[
    \bbE [\Psi^A_k - \Psi^A_{k-1}] \ge 1 - 6\epsilon - 3\alpha_{2k-1} - 3\alpha_{2k}.
\]
We use Claim~\ref{claim:PsiA-Psi} to assist us in calculating $\bbE [ \Psi^A_k - \Psi^A_{k-1} ]$. Define $U^B_{2k-1}$ to be $1$ if Bob makes a good update, $-1$ if he makes a bad update, and $0$ if he makes no update in response to message $2k-1$ from Alice. Let $\ECC(y_{2k-1}, x_{2k-1}, \delta_{2k-1})$ be Alice's intended message for message $2k$, and let $m_{2k-1}$ be the message Bob receives. Let $(x^*_{2k-1}, y^*_{2k-1}, \delta^*_{2k-1}) \in \Sigma$ with a corresponding $p^*_{2k-1}$ be what Bob computes in his protocol, if they exist. Let $d^*_{2k-1}=d_{m_{2k-1}}(y^*_{2k-1}, x^*_{2k-1}, \delta^*_{2k-1})$. Note that the triple $(y_{2k-1}, x_{2k-1}, \delta_{2k-1})$ is not necessarily deterministic, but we will show that Equations~\eqref{eqn:alice-0.5-3alpha} and~\eqref{eqn:bob-0.5-3alpha} hold for any specific value of $(y_{2k-1}, x_{2k-1}, \delta_{2k-1})$.

The strategy is to note that 
\begin{align*}
    \bbE[\Psi^A_{k} - \Psi^A_{k-1}] 
    &= \bbE[(\Psi_k - \Psi^A_{k-1}) + (\Psi^A_k - \Psi_k)].
\end{align*}
We have already calculated $\bbE[\Psi^A_k - \Psi_k]$ in Claim~\ref{claim:PsiA-Psi}, so it remains to analyze the term $\Psi_k - \Psi^A_{k-1}$. 

We split the analysis into two cases based on the value of $\psi^B_{2k-2}$. 

\paragraph{Proof of Lemma~\ref{lemma:0.5-3alpha} when $\psi^B_{2k-2} < n$.} In this case, we have that
\begin{align*}
    \Psi_k - \Psi^A_{k-1} 
    &= \psi^A_{2k-1} + \rho^A_{2k} + \min \{ \psi^B_{2k-1}, n \} - \psi^A_{2k-2} - \min \{ \psi^B_{2k-2} + \min \{ \rho^B_{2k-1}, 0.5 \}, n \} \\
    &= \rho^A_{2k} + U^B_{2k-1} - \min\{ \rho^B_{2k-1}, 0.5 \},
\end{align*}
since $\psi^A_{2k-1} = \psi^A_{2k-2}$ and $\rho_{2k-2} + \min \{ \rho^B_{2k-1}, 0.5 \}, \rho^B_{2k-1} \le n$. It thus holds that
\begin{align*}
    \bbE[\Psi^A_{k} - \Psi^A_{k-1}] 
    &= \bbE[(\rho^A_{2k} + U^B_{2k-1} - \min\{ \rho^B_{2k-1}, 0.5 \}) + (\Psi^A_k - \Psi_k)] \\
    &= \bbE[U^B_{2k-1} - \min\{ \rho^B_{2k-1}, 0.5 \} + \rho^A_{2k} + (\Psi^A_k - \Psi_k)].
\end{align*}
Again, we split the analysis into cases. 

\begin{caseof}
\begin{mdframed}[topline=false,rightline=false,bottomline=false]
\case {$(y^*_{2k-1},x^*_{2k-1},\delta^*_{2k-1})$ does not exist.}{
\vspace{0.075in}
\begin{itemize}
    \item $\bbE[U^B_{2k-1}]=0$ because Bob does not perform an update.
    \item $\rho^A_{2k} \ge 0.5$ because Bob's next message is $\ECC(\hat{x}, y, \bullet)$.
    \item $\bbE[\Psi^A_k - \Psi_k] \ge 0.5 - 3\epsilon - 3\alpha_{2k-1}$ since $\psi^B_{2k-1} = \psi^B_{2k-2} < n$ because Bob never updates.
    \item $\alpha_{2k-1}\geq \left(\frac16-\epsilon\right)$ because $(y^*_{2k-1},x^*_{2k-1},\delta^*_{2k-1})$ does not exist.
\end{itemize}
\vspace*{-\baselineskip}
    \begin{align*}
        \implies \bbE[\Psi^A_k - \Psi^A_{k-1}] 
        &\ge \bbE[U^B_{2k-1}] - \min \{ \rho^B_{2k-1}, 0.5 \} + \bbE[\rho^A_{2k}] + \bbE[(\Psi^A_k - \Psi_k)] \\
        &\ge 0 - 0.5 + 0.5 + (0.5 - 3\epsilon - 3\alpha_{2k}) \\
        &\ge 1 - 6\epsilon - 3\alpha_{2k-1} - 3\alpha_{2k}.
    \end{align*}
}
\case{$(y_{2k-1},x_{2k-1},\delta_{2k-1}) = (y^*_{2k-1},x^*_{2k-1},\delta^*_{2k-1})$ and ($x_{2k-1} = \hat{x}$ or $\delta_{2k-1} = \bullet$).}{ 
\vspace{0.075in}
\begin{itemize}
    \item $\bbE[U^B_{2k-1}] \ge 0.5 - 3\alpha_{2k-1}$. Bob makes a good update with probability $p^*_{2k-1}\ge0.5-3d^*_{2k-1}$ and $\alpha_{2k-1}\geq d^*_{2k-1}$ giving the desired result.
    \item $\bbE[\rho^A_{2k} + (\Psi^A_k - \Psi_k)] \ge 1 - 3\epsilon - 3\alpha_{2k}$. To see this, we consider if $\psi^B_{2k-1} \ge n$ or $\psi^B_{2k-1} < n$. If $\psi^B_{2k-1} \ge n$, Bob must've made a good update to message $2k-1$, so it holds that $\rho^A_{2k} = 1$ (since Bob's next message is $\ECC(x, y_{2k-1} = \hat{y}, \op_y(y_{2k-1}))$) and $\bbE[\Psi^A_k - \Psi_k ~|~ \psi^B_{2k-1} \ge n] \ge -3\epsilon-3\alpha_{2k}$, so that $\bbE[\rho^A_{2k} + (\Psi^A_k - \Psi_k)~|~ \psi^B_{2k-1} \ge n] \ge 1 -3\epsilon- 3\alpha_{2k}$. 
    
    On the other hand, if $\psi^B_{2k-1} < n$, $\rho^A_{2k} \ge 0.5$ since Bob's next message is $\ECC(\hat{x}, y_{2k-1}, \op_y(y_{2k-1}))$, and $\bbE[\Psi^A_k - \Psi_k ~|~ \psi^B_{2k-1} < n] \ge 0.5 - 3\epsilon - 3\alpha_{2k}$, so that $\bbE[\rho^A_{2k} + (\Psi^A_k - \Psi_k) ~|~ \psi^B_{2k-1} < n] \ge 1 - 3\epsilon - 3\alpha_{2k}$. 
\end{itemize}
\vspace*{-\baselineskip}
    \begin{align*}
        \implies \bbE[\Psi^A_k - \Psi^A_{k-1}] 
        &\ge \bbE[U^B_{2k-1}] - \min \{ \rho^B_{2k-1}, 0.5 \} + \bbE[\rho^A_{2k} + (\Psi^A_k - \Psi_k)] \\
        &\ge (0.5 - 3\alpha_{2k-1}) - 0.5 + (1 - 3\epsilon - 3\alpha_{2k}) \\
        &= 1 - 6\epsilon - 3\alpha_{2k-1} - 3\alpha_{2k}.
    \end{align*}
} 
\case {$(y_{2k-1},x_{2k-1},\delta_{2k-1}) = (y^*_{2k-1},x^*_{2k-1},\delta^*_{2k-1})$ and ($x_{2k-1}\not=\hat{x}$ and $\delta_{2k-1} \not= \bullet$).}{
\vspace{0.075in}
\begin{itemize}
    \item $U^B_{2k-1} = 0$ because Bob does not perform an update for the given values of $x^*_{2k-1}$ and $\delta^*_{2k-1}$.
    \item $\rho^B_{2k-1} = 0$ for the same reason.
    \item $\rho^A_{2k} \ge 0.5$ because Bob's next message is $\ECC(\hat{x}, y_{2k-1}, \op_y(y_{2k-1}))$.
    \item $\bbE[\Psi^A_k - \Psi_k] \ge 0.5-3\epsilon - 3\alpha_{2k}$ since $\psi^B_{2k-1} = \psi^B_{2k-2} < n$.
\end{itemize}
\vspace*{-\baselineskip}
    \begin{align*}
        \implies \bbE[\Psi^A_k - \Psi^A_{k-1}] 
        &\ge \bbE[U^B_{2k-1}] - \min \{ \rho^B_{2k-1}, 0.5 \} + \bbE[\rho^A_{2k}] + \bbE[(\Psi^A_k - \Psi_k)] \\
        &\ge 0 - 0 + 0.5 + (0.5 - 3\epsilon - 3\alpha_{2k}) \\
        &\ge 1 - 6\epsilon - 3\alpha_{2k-1} - 3\alpha_{2k}.
    \end{align*}
}
\case {$(y_{2k-1},x_{2k-1},\delta_{2k-1}) \neq (y^*_{2k-1},x^*_{2k-1},\delta^*_{2k-1})$ and ($y^*_{2k-1} = y$ and $x^*_{2k-1} = \hat{x}$).}{
\begin{subcaseof}
    \subcase{The update corresponding to Bob receiving $\ECC(y^*_{2k-1}, x^*_{2k-1}, \delta^*_{2k-1})$ is bad or none.}{
    \begin{itemize}
        \item $\bbE[U^B_{2k-1}] \ge 0.5 - 3\epsilon - 3\alpha_{2k-1}$. Bob makes a bad update with probability at most $p_{2k-1} \le 1 - 3d^*_{2k-1}$ and $d^*_{2k-1} \ge (\frac12 - \epsilon) - \alpha_{2k-1}$.
        \item $\rho^A_{2k} \ge 0.5$ because Bob's response is $\ECC(\hat{x}, y, \bullet)$
        \item $\bbE[\Psi^A_k - \Psi_k] \ge 0.5-3\epsilon - 3\alpha_{2k}$ because Bob could only have made a bad update, so $\psi^B_{2k-1}<n$.
    \end{itemize}
    \vspace*{-\baselineskip}
    \begin{align*}
        \implies \bbE[\Psi^A_k - \Psi^A_{k-1}] 
        &\ge \bbE[U^B_{2k-1}] - \min \{ \rho^B_{2k-1}, 0.5 \} + \bbE[\rho^A_{2k}] + \bbE[\Psi^A_k - \Psi_k] \\
        &\ge (0.5 - 3\epsilon - 3\alpha_{2k-1}) - 0.5 + 0.5 + (0.5 - 3\epsilon - 3\alpha_{2k}) \\
        &= 1 - 6\epsilon - 3\alpha_{2k-1} - 3\alpha_{2k}.
    \end{align*}
    }
    \subcase{The update corresponding to Bob receiving $\ECC(y^*_{2k-1}, x^*_{2k-1}, \delta^*_{2k-1})$ is good.}{
    \begin{itemize}
        \item $U^B_{2k-1} + \bbE[\Psi^A_k - \Psi_k] \ge 0.5 - 3\epsilon - 3\alpha_{2k}$. In the case that $\psi^B_{2k-1} \ge n$, Bob must've made a good update to message $2k-1$, so $U^B_{2k-1} = 1$, and we have that $\bbE[\Psi^A_k - \Psi_k] \ge -3\epsilon-3\alpha_{2k}$. In the case that $\psi^B_{2k-1} < n$, we have $U^B_{2k-1} \ge 0$ and $\bbE[\Psi^A_k - \Psi_k] \ge 0.5-3\epsilon-3\alpha_{2k}$.
        \item $\rho^A_{2k} \ge 0.5$ because Bob's response is $\ECC(\hat{x}, y, \bullet)$.
        \item $\alpha_{2k-1} \ge \left(\frac16-\epsilon\right)$ because $(y_{2k-1},x_{2k-1},\delta_{2k-1}) \neq (y^*_{2k-1},x^*_{2k-1},\delta^*_{2k-1})$ so $\alpha_{2k-1} \ge (\frac12 - \epsilon) - d^*_{2k-1} \ge \frac12 - \epsilon - \frac13 = \frac16 - \epsilon$.
    \end{itemize}
    \vspace*{-\baselineskip}
        \begin{align*}
            \implies \bbE[\Psi^A_k - \Psi^A_{k-1}] 
            &\ge \bbE[U^B_{2k-1} + (\Psi^A_k - \Psi_k)] - \min \{ \rho^B_{2k-1}, 0.5 \} + \bbE[\rho^A_{2k}] \\
            &\ge (0.5 - 3\epsilon - 3\alpha_{2k}) - 0.5 + 0.5 \\
            &\ge 1 - 6\epsilon - 3\alpha_{2k-1} - 3\alpha_{2k},
        \end{align*}
    }
\end{subcaseof}
}
\case {$(y_{2k-1},x_{2k-1},\delta_{2k-1}) \neq (y^*_{2k-1},x^*_{2k-1},\delta^*_{2k-1})$ and ($y^*_{2k-1} \not= y$ or $x^*_{2k-1} \not= \hat{x}$).}{
\begin{subcaseof}
    \subcase{The update corresponding to Bob receiving $\ECC(y^*_{2k-1}, x^*_{2k-1}, \delta^*_{2k-1})$ is bad or none.}{
    \begin{itemize}
        \item $\bbE[U^B_{2k-1}] \ge 1 - 3\epsilon - 3\alpha_{2k-1}$ because Bob makes a bad update with probability at most $p_{2k-1} \le 0.5 - 3d^*_{2k-1}$ and $d^*_{2k-1} \geq \left(\frac12 - \epsilon\right) - \alpha_{2k-1}$.
        \item $\bbE[\Psi^A_k - \Psi_k] \ge 0.5 - 3\epsilon - 3\alpha_{2k}$ because Bob could only have made a bad update so $\psi^B_{2k-1} < n$.
    \end{itemize}
    \vspace*{-\baselineskip}
    \begin{align*}
        \implies \bbE[\Psi^A_k - \Psi^A_{k-1}] 
        &\ge \bbE[U^B_{2k-1}] - \min \{ \rho^B_{2k-1}, 0.5 \} + \bbE[\rho^A_{2k}] + \bbE[\Psi^A_k - \Psi_k] \\
        &\ge (1 - 3\epsilon - 3\alpha_{2k-1})  - 0.5 + 0 + (0.5 - 3\epsilon - 3\alpha_{2k}) \\
        &= 1 - 6\epsilon - 3\alpha_{2k-1} - 3\alpha_{2k}.
    \end{align*}
    }
    \subcase{The update corresponding to Bob receiving $\ECC(y^*_{2k-1}, x^*_{2k-1}, \delta^*_{2k-1})$ is good.}{
    \begin{itemize}
        \item $U^B_{2k-1} + \bbE[\Psi^A_k - \Psi_k] \ge 0.5 - 3\epsilon - 3\alpha_{2k}$. If $U^B_{2k-1} = 1$, then $\bbE[\Psi^A_k - \Psi_k] \ge -3\epsilon - 3\alpha_{2k}$. If $U^B_{2k-1} = 0$, then $\bbE[\Psi^A_k - \Psi_k] \ge 0.5 - 3\epsilon - 3\alpha_{2k}$. Either way $U^B_{2k-1} + \bbE[\Psi^A_k - \Psi_k] \ge 0.5 - 3\epsilon - 3\alpha_{2k}$.
        
        \item $\alpha_{2k-1} \ge \frac13$ because $(y_{2k-1},x_{2k-1},\delta_{2k-1}) \neq (y^*_{2k-1},x^*_{2k-1},\delta^*_{2k-1})$ so $\alpha_{2k-1} \ge \left(\frac12 - \epsilon\right) - d^*_{2k-1} \ge \left(\frac12 - \epsilon\right) - \left(\frac16 - \epsilon\right) = \frac13$.
    \end{itemize}
    \vspace*{-\baselineskip}
    \begin{align*}
        \implies \bbE[\Psi^A_k - \Psi^A_{k-1}] 
        &\ge \bbE[U^B_{2k-1} + (\Psi^A_k - \Psi_k)] - \min \{ \rho^B_{2k-1}, 0.5 \} + \rho^A_{2k} \\
        &\ge (0.5 - 3\epsilon - 3\alpha_{2k}) - 0.5 + 0 \\
        &\ge 1 - 6\epsilon - 3\alpha_{2k-1} - 3\alpha_{2k},
    \end{align*}
    }
\end{subcaseof}
}
\end{mdframed}
\end{caseof}

\noindent
\paragraph{Proof of Lemma~\ref{lemma:0.5-3alpha} when $\psi^B_{2k-2} \geq n$.} Finally, we consider the case where $\psi^B_{2k-2} \ge n$. We have that
\begin{align*}
    \Psi_k - \Psi^A_{k-1} 
    &= \psi^A_{2k-1} + \rho^A_{2k} + \min \{ \psi^B_{2k-1}, n \} - \psi^A_{2k-2} - \min \{ \psi^B_{2k-2} + \min \{ \rho^B_{2k-1}, 0.5 \}, n \} \\
    &= \rho^A_{2k} + \min \{ \psi^B_{2k-1}, n \} - n
\end{align*}
since $\psi^A_{2k-1} = \psi^A_{2k-2}$ and $\psi^B_{2k-2} + \min \{ \rho^B_{2k-1}, 0.5 \} \ge \psi^B_{2k-2} \ge n$. Then
\begin{align*}
    \bbE[\Psi^A_{k} - \Psi^A_{k-1}] 
    &= \bbE[\rho^A_{2k} + (\min \{ \psi^B_{2k-1}, n \} - n) + (\Psi^A_k - \Psi_k)].
\end{align*}

Again, we split the analysis into cases. Note that $\bbE[\Psi^A_k - \Psi_k]\geq -3\epsilon-3\alpha_{2k}$ by Claim~\ref{claim:PsiA-Psi}.

\noindent
\begin{caseof}
\begin{mdframed}[topline=false,rightline=false,bottomline=false]
\case {$(y^*_{2k-1},x^*_{2k-1},\delta^*_{2k-1})$ does not exist.}{
\vspace{0.075in}
\begin{itemize}
    \item $\rho^A_{2k} \ge 0.5$ because Bob's next message is $\ECC(\hat{x}, y, \bullet)$. 
    \item $\psi^B_{2k-1} \ge n$ still because Bob does not update.
    \item $\alpha_{2k-1} \geq \left(\frac16-\epsilon\right)$ becuase $(y^*_{2k-1},x^*_{2k-1},\delta^*_{2k-1})$ does not exist.
    
\end{itemize}
    \vspace*{-\baselineskip}
    \begin{align*}
        \implies \bbE[\Psi^A_k - \Psi^A_{k-1}] 
        &= \bbE[\rho^A_{2k} + \min \{ \psi^B_{2k-1}, n \} - n] + \bbE[\Psi^A_k - \Psi_k] \\
        &\ge 0.5 + n - n + (- 3\epsilon - 3\alpha_{2k}) \\
        &\ge 1 - 6\epsilon - 3\alpha_{2k-1} - 3\alpha_{2k}.
    \end{align*}
}
\case{$(y_{2k-1},x_{2k-1},\delta_{2k-1}) = (y^*_{2k-1},x^*_{2k-1},\delta^*_{2k-1})$.}{ 
\vspace{0.075in}
\begin{itemize}
    \item $\psi^B_{2k-1} \ge n$ because Bob cannot perform a bad update.
    \item $\rho^A_{2k} = 1$ because Bob's next message is $\ECC(\hat{x} = x, y_{2k-1} = \hat{y}, \op_y(y_{2k-1}))$.
\end{itemize}
    \vspace*{-\baselineskip}
    \begin{align*}
        \implies \bbE[\Psi^A_k - \Psi^A_{k-1}] 
        &= \bbE[\rho^A_{2k} + \min \{ \psi^B_{2k-1}, n \} - n] + \bbE[\Psi^A_k - \Psi_k] \\
        &\ge 1 + n - n + (- 3\epsilon - 3\alpha_{2k}) \\
        &\ge 1 - 6\epsilon - 3\alpha_{2k-1} - 3\alpha_{2k}.
    \end{align*}
}
\case{$(y_{2k-1},x_{2k-1},\delta_{2k-1}) \neq (y^*_{2k-1},x^*_{2k-1},\delta^*_{2k-1})$ and ($y^*_{2k-1}=y$ and $x^*_{2k-1}=\hat{x}$).}{ 
\vspace{0.075in}
\begin{itemize}
    \item $\rho^A_{2k} \ge 0.5$ because Bob's next message to Alice is $\ECC(\hat{x}, y, \bullet)$.
    \item $\bbE [ \min \{ \psi^B_{2k-1}, n \} - n ] \ge 0.5 - 3\epsilon - 3\alpha_{2k-1}$. Bob makes a bad update with probability $\le p^*_{2k-1} = 1 - 3d^*_{2k-1}$. Then $\bbE [ \min \{ \psi^B_{2k-1}, n \} - n ] \ge 3d^*_{2k-1}-1$, so using $d^*_{2k-1}\geq \left(\frac12-\epsilon\right)-\alpha_{2k-1}$ gives the desired result.
\end{itemize}
\vspace*{-\baselineskip}
    \begin{align*}
        \implies \bbE[\Psi^A_k - \Psi^A_{k-1}] 
        &= \bbE[\rho^A_{2k} + \min \{ \psi^B_{2k-1}, n \} - n] + \bbE[\Psi^A_k - \Psi_k] \\
        &\ge 0.5 + (0.5 - 3\epsilon - 3\alpha_{2k-1}) + (- 3\epsilon - 3\alpha_{2k}) \\
        &= 1 - 6\epsilon - 3\alpha_{2k-1} - 3\alpha_{2k}.
    \end{align*}
}
\case{$(y_{2k-1},x_{2k-1},\delta_{2k-1}) \neq (y^*_{2k-1},x^*_{2k-1},\delta^*_{2k-1})$ and ($y^*_{2k-1}\neq y$ or $x^*_{2k-1}\neq \hat{x}$).}{ 
\vspace{0.075in}
\begin{itemize}
    \item $\rho^A_{2k} \ge 0$ because Alice will never send a message causing Bob to make a bad update.
    \item $\bbE[\min \{ \psi^B_{2k-1}, n \} - n] \ge 1 - 3\epsilon - 3\alpha_{2k-1}$. Bob makes a bad update with probability $\le p^*_{2k-1} = 0.5 - 3d^*_{2k-1}$. Using $d^*_{2k-1}\geq \left(\frac12-\epsilon\right)-\alpha_{2k-1}$ gives $\bbE[\min \{ \psi^B_{2k-1}, n \} - n] \ge -(0.5 - 3d^*_{2k-1}) \ge 1 - 3\epsilon - 3\alpha_{2k-1}$. 
\end{itemize}
\vspace*{-\baselineskip}
    \begin{align*}
        \implies\bbE[\Psi^A_k - \Psi^A_{k-1}] 
        &= \bbE[\rho^A_{2k}] + \bbE[\min \{ \psi^B_{2k-1}, n \} - n] + \bbE[\Psi^A_k - \Psi_k] \\
        &\ge 0 + (1 - 3\epsilon - 3\alpha_{2k-1}) + (- 3\epsilon - 3\alpha_{2k}) \\
        &= 1 - 6\epsilon - 3\alpha_{2k-1} - 3\alpha_{2k}.
    \end{align*}
}
\end{mdframed}
\end{caseof}




\section{Binary Interactive Protocol Resilient to $\frac12$ Erasures} \label{sec:erasure}

In this section, we present a non-adaptive interactive protocol where Alice and Bob simulate an existing protocol $\pi_0$ via a protocol $\pi$ resilient to $\frac12-\epsilon$ erasures, for any $\epsilon>0$.

This result is optimal; no protocol is resilient to $\frac12$ erasures for all possible functions Alice and Bob might want to compute.

\begin{theorem} [\cite{FranklinGOS15}]
    For all $n>0$, there exists a function $f(x, y)$ of Alice and Bob's inputs $x,y\in\{0,1\}^n$, such that any non-adaptive interactive protocol over the binary erasure channel that computes $f(x, y)$ succeeds with probability at most $\frac12$ if a $\frac12$ fraction of the transmissions are erased.
\end{theorem}

In our protocol, Alice and Bob jointly recreate the transcript of $\pi_0$ by sending their next message in $\pi_0$ until they are sure the other party has received it. How they do this while staying in sync is explained informally in Section~\ref{sec:overview-1/2}.

\subsection{Formal Protocol}

\protocol{Protocol Resilient to $\frac12 - 4\epsilon$ Erasures}{1/2-erasure}{
    Let $\pi_0$ be an error-free binary protocol that runs in $n_0$ rounds. We may assume that $\pi_0$ is alternating (i.e. Alice and Bob take turns speaking with Alice speaking first) with at most a factor of $2$ blowup in the round complexity. We pad $\pi_0$ with $1$'s so that past round $n_0$, Alice and Bob both send $1$ every round. 
    
    \vspace{0.075in}
    
    Our protocol $\pi$ occurs in $N = \frac{2n_0}{\epsilon^2}$ rounds, where Alice and Bob alternate speaking with Alice speaking first. These rounds are partitioned into $\frac{n_0}{\epsilon}$ blocks of $\frac{2}{\epsilon}$ rounds, so that in each block Alice and Bob each speak $\frac1\epsilon$ bits alternatingly.\footnote{We can assume that $\frac{1}{\epsilon}$ is an integer; otherwise, take a slightly smaller $\epsilon$ for which $\frac{1}{\epsilon} \in \bbN$.}
    
    \vspace{0.075in}
    
    Alice and Bob each have an internal mode, which is either $\speaker$, $\listener$, or $\passer$. They also track an internal transcript $T$ equal to what they believe the current noiseless protocol is, including the next bit they are trying to send. Thus, initially, $T = \emptyset$ for Bob, and for Alice $T$ is her first message of $\pi$. Our protocol begins with Alice in $\speaker$ mode and Bob in $\listener$ mode.
    
    \vspace{0.075in}
    
    Alice and Bob stay in the same mode for an entire block, only potentially transitioning at the end of the block. Their behavior in each mode is described as follows. 
    
    \begin{enumerate}[align=left, leftmargin=*, label={\bf zzzzzz:}]
    
    \item[$\listener$:] 
        \begin{itemize}[leftmargin=*]
            \item Let $\beta \in \{ 0, 1, \perp \}$ be the most recently received message from the other party. If $\beta = \perp$, send $0$. Otherwise send $1$.
            \item \emph{Transition Condition:} If the first non-erased bit received in the block was $0$, switch to being in $\speaker$ mode at the end of the block. Also recall the sequence of bits received in the last block prior to this one where not all incoming messages were erased, and let $b = 0$ if at least one $0$ was received and $b = 1$ otherwise. Let $b'$ be the next bit the party would send if the transcript so far is $T || b$, and set $T \gets T || b || b'$. 
        \end{itemize}
        
    \item[$\speaker$:] 
        \begin{itemize}[leftmargin=*]
            \item Let $b$ be the the last bit of the the party's transcript $T$ ($b$ is their next message to send). Send $1$ repeatedly until receiving a $1$, then send $b$ for the rest of the block.
            \item \emph{Transition Condition:} If ever $1$ was received after having switched to sending $b$, switch to being in $\passer$ mode at the end of the block. Otherwise, remain in $\speaker$ mode for the next block.
        \end{itemize}
        
    \item[$\passer$:] 
        \begin{itemize}[leftmargin=*]
            \item Always send $0$.
            \item \emph{Transition Condition:} If a $1$ was received in the block, switch to being in $\listener$ mode at the end of the block.
        \end{itemize}
    \end{enumerate}
    
}
\subsection{Analysis}

We begin by proving several claims about Protocol~\ref{prot:1/2-erasure}.

\begin{claim} \label{claim:modes}
    Alice and Bob are never in the same mode at the same time.
\end{claim}

\begin{proof}
    We show that if Alice and Bob are in two different modes at the beginning of a block, then they cannot be in the same mode at the end of that block. Note that they transition between modes according to the cycle $\listener \to \speaker \to \passer \to \cdots$ and cannot transition more than once per block.
    
    First assume Alice and Bob are $\listener$ and $\speaker$ in some order at the beginning of a block. It suffices to show the $\listener$ cannot become a $\speaker$ within that block. This is true because the $\listener$ only transitions if the first bit heard within the block is a $0$ but the $\speaker$ only sends $0$ after they receive a $1$ confirming that a $1$ has been received.
    
    If Alice and Bob are $\passer$ and $\listener$ in some order, we show that the $\passer$ can only become a $\listener$ if the $\listener$ becomes a $\speaker$. The $\passer$ only becomes a $\listener$ when they receive a $1$, and the $\listener$ only sends $1$ when they receive a (non-erased) bit from the $\passer$. This bit received from the $\passer$ must be a $0$, which will cause the $\listener$ to transition to $\speaker$ mode.
    
    Finally, if Alice and Bob are $\speaker$ and $\passer$ in some order it suffices to show the $\speaker$ cannot become a $\passer$. This is true because the $\passer$ only ever sends $0$ in the block, and the $\speaker$ only transitions if they heard at least two $1$'s.
\end{proof}

\begin{claim}
    The party that most recently left $\listener$ mode (or Alice if it is the start of the protocol) has a transcript $T$ that is $1$ bit longer than the other party's.
\end{claim}

\begin{proof}
    This is true at the start of the protocol: Alice's $T$ is length $1$, and Bob's $T$ is length $0$. Since Alice and Bob cycle through $\listener \to \speaker \to \passer \to \cdots$ without ever being in the same mode at the same time by Claim~\ref{claim:modes}, they alternate leaving $\listener$ mode starting with Bob (who begins in $\listener$ mode). Thus, every time a party leaves $\listener$ mode they add $2$ bits to $T$, and the claim follows.
\end{proof}

\begin{claim} \label{claim:transcript}
    On inputs $x,y$, a party's simulated transcript $T$ is always a prefix of $\pi_0(x,y)$.
\end{claim}

\begin{proof}
    A party only modifies $T$ upon exiting $\listener$ mode, when they add two bits, one for the other party's message and one for their own. We show that the first bit they added must be the correct next bit of $\pi_0(x, y)$; it then follows that both bits must be the correct next two bits.
    
    In the block $B$ before the party (w.l.o.g. Alice) exits $\listener$ mode, the first bit she received in that block must've been a $0$. This implies that Bob was in $\passer$ mode in block $B$: he cannot also be in $\listener$ mode by Claim~\ref{claim:modes}, and he cannot be in $\speaker$ mode since then he'd only send $0$ \emph{after} receiving a $1$ confirming Alice's reception of a $1$. The last block $R$ prior to $B$ that Bob was in $\speaker$ mode trying to send a bit $b$, he received a $1$ from Alice confirming the reception of the bit $b$. Then, Alice must've received a nonzero sequence of $1$'s followed by a nonzero sequence of $b$'s in block $R$. This must have also been the last block prior to block $B$ that Alice received \emph{any} bits: Bob sent only $0$'s in blocks $R+1, \dots, B$, and block $B$ is the first time that Alice received a $0$. Thus, Alice determines Bob's bit $b$ correctly and appends it and her next message to her transcript.
    
    
\end{proof}
    
\begin{claim}
    If a block has at most $\frac{1}{\epsilon}-3$ corruptions, then at least one of Alice and Bob transitions modes at the end of the block. 
\end{claim}

\begin{proof}
    To show this, we consider the possible combinations of Alice and Bob's starting modes. 
    
    If Alice and Bob are in $\speaker$ and $\listener$ mode, respectively, there are at least $2$ (in fact $3$) pairs of consecutive Alice-Bob rounds for which neither message is erased, since the adversary can only erase half of the communication for all but $3$ Alice-Bob pairs. Let the bit of $\pi_0$ Alice is trying to send be $b$. Then, in the first such pair, Alice sends $1$ and receives a $1$ from Bob, and in the second such pair, she sends $b$ and receives a $1$. Then, at the end of the block, she transitions to $\passer$ mode. The case where Alice is $\listener$ and Bob is $\speaker$ is identical, except we disregard Alice's first and last rounds and consider Bob-Alice pairs of messages. There are still at least $2$ non-erased pairs, which is enough for Bob to communicate the two bits $1, b$ and receive confirmation bits.
    
    If Alice and Bob are in $\passer$ and $\listener$ mode, respectively, then consider Alice-Bob pairs of consecutive rounds. There is at least $1$ such pair with no erasures. In this pair, Bob must hear Alice's $0$ so that he leaves $\listener$ mode at the end of the block. The case where Alice is in $\listener$ mode and Bob is in $\passer$ mode works analogously by grouping Bob-Alice rounds, ignoring the first and last rounds of the block.
    
    If Alice and Bob are in $\speaker$ and $\passer$ mode, respectively, Bob only sends $0$'s so that Alice only sends $1$'s the entire block. Consider Alice-Bob pairs of consecutive rounds. There is at least $1$ such pair with no erasures. In the first such pair, Bob hears Alice's $1$ so that he switches to $\listener$ mode at the end of the block. The case where Alice is in $\passer$ mode and Bob is in $\speaker$ mode works analogously, group Bob-Alice rounds and ignoring the first and last rounds of the block.
\end{proof}
    
\begin{theorem}
    Protocol~\ref{prot:1/2-erasure} is resilient to $\frac12 - 4\epsilon$ fraction of erasures.
\end{theorem}

\begin{proof}
    First we claim that if there are at least $3n_0 + 6$ blocks at the end of which someone switches modes, then each party must leave $\listener$ mode at least $\frac{n_0}{2}$ times. To see this, note that at least one party must've switched modes at least $\frac{3n_0}{2} + 3$ times, so that they cycled through all three modes at least $\frac{n_0}{2} + 1$ times. Since Alice and Bob are never in the same mode at the same time, this implies that the other party must've cycled through all three modes at least $\frac{n_0}{2}$ times. In particular, both parties left $\listener$ mode at least $\frac{n_0}{2}$ times.
    
    
    Each time a party leaves $\listener$ mode, their transcript increases by length $2$, so each party has a final transcript length of at least $n_0$. By Claim~\ref{claim:transcript} this final transcript is correct.

    Now suppose that there are $\le \left( \frac12 - 4\epsilon \right)$ total erasures. Let $\hat{n}$ denote the number of blocks with at most $\frac1\epsilon - 3$ erasures. Then $\hat{n}$ satisfies the following inequality double counting the total number of erasures: 
    \begin{align*}
        & \left( \frac12 - 4\epsilon \right) \cdot \frac{2n_0}{\epsilon^2} \ge \hat{n} \cdot 0 + \left(\frac{n_0}{\epsilon} - \hat{n} \right) \cdot \left(  \frac{1}{\epsilon}-2 \right) \\
        &\implies \frac{\hat{n}}\epsilon > \left( \frac1\epsilon - 2 \right) \cdot \hat{n} \ge \frac{6n_0}{\epsilon} \\
        &\implies \hat{n} > 6n_0 \ge 3n_0 + 3\cdot 2 = 3n_0 + 6.
    \end{align*}
    In the last step, we can assume $n_0\geq 2$ because Alice and Bob talk at least once each in $\pi_0$. As such, the number of blocks where someone switches modes is at least $3n_0+6$, so Alice and Bob must both have a correct final transcript of length at least $n_0$ at the end of the protocol.
    
\end{proof}
\section{Acknowledgments}

The authors would like to thank their advisor, Dr. Yael Tauman Kalai (Microsoft Research and MIT). They would like to thank her for helpful discussions, paper edits, as well as introducing them to interactive coding and others who have studied it in the past. They would also like to thank Michael Kural and Naren Manoj for reading the paper and providing comments.

Rachel Yun Zhang is supported by an Akamai Presidential Fellowship.

\bibliographystyle{alpha}
\bibliography{refs}

\end{document}